\RequirePackage{fix-cm}
\documentclass[smallextended]{svjour3}       
\smartqed  
\usepackage{graphicx}
\usepackage{amsmath}
\usepackage{amssymb}

\usepackage{enumerate}
\usepackage{hhline}

\newcommand{\ignore}[1]{}

\newcommand{\boxtheorem}{\hfill $\Box$}
\newcommand{\nit}[1]{{\it #1}}

\newcommand{\IC}{\nit{IC}}

\newcommand{\sem}{{\succ,\mc{X}\!}}
\newcommand{\nn}{\nit{null}}

\newcommand{\semc}{{\succ_{\!c},\mc{X}\!}}
\newcommand{\semcr}{{\succ_{\!c}^r,\mc{X}\!}}

\newcommand{\red}{}

\newcommand{\comlb}[1]{{\vspace{2mm}\noindent \red{\bf COMM(LEO):}}~ #1 \hfill {\bf    END.}\\}
\newcommand{\combabak}[1]{{\vspace{4mm}\noindent \bf  COMM(Babak):}~ \red {\em  #1}\hfill {\bf END.}\\}
\newcommand{\mc}[1]{\mathcal{ #1}}
\newcommand{\mf}[1]{\mathfrak{ #1}}

\newcommand{\defproof}[2]{{\noindent\bf Proof of #1:\
}#2 \boxtheorem\\}

\begin{document}

\title{From Causes for Database Queries to Repairs and Model-Based Diagnosis and  Back
}

\titlerunning{From Causes for Database Queries}        

\author{Leopoldo Bertossi \and  Babak Salimi
}


\institute{L. Bertossi \at
           Carleton University \\
              School of Computer Science\\
              Ottawa, Canada.\\
              \email{bertossi@scs.carleton.ca}
              \and
              B. Salimi \at
             University of Washington \\
              Computer Science and Engineering\\
              Seattle, USA.\\
              \email{bsalimi@cs.washington.edu}
           }


\date{}

\maketitle

\begin{abstract}
In this work we establish and investigate connections between causes for query answers in databases, database repairs with respect to denial constraints, and consistency-based diagnosis.
The first two are relatively new research areas in databases, and the third one is an established subject in knowledge representation. We show how to obtain database repairs from causes, and
the other way around. Causality problems are formulated as diagnosis problems, and the diagnoses provide causes and their responsibilities. The vast
body of research on database repairs can be applied to the newer problems of computing actual causes for query answers and their responsibilities.
 These connections are interesting {\em per se}. They also allow us, after a transition inspired by consistency-based diagnosis
 to computational problems on hitting-sets and vertex covers
 in hypergraphs,  to obtain several new algorithmic and complexity results for database causality.
\end{abstract}

\section{\ Introduction}
\vspace{-1mm}
When querying a database, a user may not always obtain the expected results, and the system could provide some explanations. They could be useful to further understand the data or  check if the
query is the intended one. Actually,
the notion of explanation for a query result was introduced in \cite{Meliou2010a}, on the basis of the deeper concept of {\em actual causation}.\footnote{In contrast with general causal claims, such as ``smoking causes cancer", which refer some sort of related events,  actual causation  specifies a particular instantiation of a causal relationship,  e.g., ``Joe's smoking is a  cause for his cancer".}

A tuple $t$  is an {\em actual cause} for an answer $\bar{a}$ to a
conjunctive query $\mc{Q}$ from  a relational database instance $D$ if there is a {\em contingent} set of tuples $\Gamma$,
such that, after removing $\Gamma$ from $D$, $\bar{a}$ is still an answer, but after further  removing $t$ from $D\smallsetminus \Gamma$, $\bar{a}$ is not an answer anymore
\red{(cf. Section \ref{causeandres} for a precise definition)}.
\red{Here, $\Gamma$ is a set of tuples that has to accompany $t$ so that the latter becomes a  counterfactual cause for answer $\bar{a}$.}
Actual causes and contingent tuples  are restricted to be among a pre-specified set
of {\em endogenous tuples}, which are admissible, possible candidates for causes, as opposed to {\em exogenous tuples}, which may also be present in the database. In rest of this paper, whenever we simply say ``cause", we mean ``actual cause".

In applications involving large
data sets, it is crucial to rank potential causes by their {\em responsibilities} \cite{Meliou2010b,Meliou2010a}, which reflect the relative (quantitative) degrees of their causality
for a query result. The responsibility measure for a cause is based on its {\em contingency sets}: the smallest (one of) its contingency sets, the strongest it is as a cause.

Actual causation, as used in  \cite{Meliou2010a}, can be traced back to
\cite{Halpern01,Halpern05}, which provides  a model-based account of causation on the basis of {\em counterfactual dependence}. {\em Causal responsibility} was introduced  in \cite{cs-AI-0312038}, to provide a graded,
 quantitative notion of causality when multiple causes may over-determine an outcome.

Apart from the explicit use of causality, research on explanations for query results has focused mainly, and rather implicitly, on provenance
\cite{BunemanKT01,BunemanT07,Cheney09,CuiWW00,Tannen10,Karvounarakis02,tannen}\ignore{, and
more recently, on provenance for non-answers \cite{ChapmanJ09,HuangCDN08}}.\ignore{\footnote{That is, tracing back, sometimes through the interplay of database tuple annotations, the reasons for {\em not} obtaining a possibly  expected answer to a query.}}
 A close connection between
causality and provenance has been established in \cite{Meliou2010a}.  However, causality is a more refined notion that identifies causes
for query results on the basis of  user-defined criteria, and ranks causes according to
their responsibilities \cite{Meliou2010b}. \ignore{For a formalization of non-causality-based explanations for
query answers in DL ontologies, see \cite{borgida}.  }

{\em Consistency-based diagnosis}  \cite{Reiter87}, a form of model-based diagnosis \cite[sec. 10.3]{struss}, is an area of knowledge representation. The problem here is, given the {\em specification} of a system in some logical formalism and a usually unexpected {\em observation}  about the system, to obtain {\em explanations} for the observation, in the form of a diagnosis for the unintended behavior \red{(cf. Section \ref{sec:mbd} for a precise definition)}.

In a different direction, a database instance, $D$, that is expected to satisfy certain integrity constraints may fail to do so. In this case, a {\em repair} of $D$ is a
database   $D'$ that does satisfy the integrity constraints and {\em minimally departs} from $D$. Different forms of minimality can be applied and investigated. A {\em consistent answer} to a query
from $D$ and with respect to\ the integrity constraints is a query answer that is obtained from all possible repairs, i.e. is invariant or certain under the class of repairs \red{(cf. Section \ref{sec:reps} for a precise definition)}. These notions were introduced
in \cite{pods99} (see \cite{Bertossi06,2011Bertossi} for surveys).\footnote{Although not in the context of repairs, consistency-based diagnosis has been applied
to consistency restoration of a database with respect to  integrity constraints \cite{Gertz97}.}

These three forms of reasoning, namely inferring causes from databases, consistency-based diagnosis, and consistent query answering (and repairs) are all {\em non-monotonic} \cite{Saliminmr2014}.
For example, a (most responsible) cause for a query result may not be such anymore after the database is updated. Furthermore, they all reflect some sort of
{\em uncertainty} about the information at hand. In this work we establish natural, precise, useful, and deeper connections
between these three reasoning tasks.

More precisely,  we unveil a
strong connection between computing causes and their responsibilities for conjunctive query answers, on one hand, and computing repairs in databases
 with respect to {\em denial constraints}, on the other. These computational
problems can be reduced to each other. In order to obtain repairs with respect to {\em a set} of denial constraints from causes, we investigate causes for queries
that are {\em unions of conjunctive queries}, and develop algorithms to compute  causes and responsibilities.

We show that inferring and computing actual causes  and their responsibilities in a database setting become
 diagnosis reasoning problems and tasks.  Actually,
a causality-based explanation for a conjunctive query answer can be viewed as a diagnosis, where in essence the first-order logical reconstruction of the relational database
 provides the system description  \cite{Reiter82}, and the observation is the query answer. \red{We obtain causes and their responsibilities -and as a side result, also database repairs-  from diagnosis.}

 Being the causality problems the main focus of this work, we take advantage of algorithms and complexity results both for consistency-based diagnosis on one side; and
 database repairs and consistent query answering \cite{2011Bertossi}, on another. In this way, we obtain new complexity results  for the main problems of causality, namely
 computing actual causes, determining their responsibilities, and obtaining most responsible causes; and also for their decision versions. In particular,
we obtain fixed-parameter polynomial-time algorithms for some of them.
 More precisely, our main results  are as follows: \ (the complexity results are all  in data complexity)

 \vspace{1mm}
 \noindent 1. \ \red{We characterize actual causes and most responsible actual causes for a boolean conjunctive query in terms of subset- and cardinality-repairs of the instance with respect to the
 denial constraint associated to the query (the query being the violation view of the constraint). In this way we can compute causes from repairs.}

 \red{In the other direction, we obtain repairs of databases with respect to {\em sets of denial constraints} from causes for query results.} For this,  we extend the treatment of causality to {\em unions of conjunctive queries}  (to represent multiple denial constraints).  We characterize an actual cause's
 responsibility in terms of cardinality-repairs. Along the way we provide PTIME algorithms to compute causes and their (minimal) contingency sets for unions of conjunctive queries.

\vspace{1mm}
 \noindent 2. \ \red{We reduce causes   for a boolean conjunctive query to  consistency-based diagnosis for the query being unexpectedly true according to a
 system description.} In particular, we show how to compute actual causes, their contingency sets, and responsibilities
 using the diagnosis characterization. \red{As a side result, we obtain database repairs from diagnosis.}

 Hitting-set-based algorithmic approaches to diagnosis \cite{Reiter87} inspire our algorithmic/complexity  approaches to causality.
In particular, \red{we reformulate} the causality problems as hitting-set problems and  vertex cover
 problems on hypergraphs, which allows us to apply results and techniques for the latter to causality.

\vspace{1mm}
 \noindent 3. \ \red{We obtain several new computational complexity results:}
 \begin{enumerate}[(a)]
 \item Checking minimal contingency sets can be done in PTIME.
 \item The {\em responsibility decision  problem} for conjunctive queries,  \red{which is about deciding if a tuple's responsibility is greater that a bound $v$ (that is part of
     the input) is \nit{NP}-complete.
 However, this problem becomes fixed-parameter tractable, with  the
 parameter being \ $\frac{1}{v}$.} \item  The problem of computing responsibilities of causes is $\nit{FP}^{\nit{NP(log(n)})}$-complete.
 Deciding most responsible causes is  $P^\nit{NP(log(n))}$-complete.

 \item The structure of the resulting hitting-set problem allows us to obtain efficient
 parameterized algorithms and
 good approximation algorithms for computing causes and minimal contingency sets.

\ignore{ \red{ \noindent 7. \ On the basis of the causality/repair connection, and the dichotomy result for causality \cite{Meliou2010a}, we obtain a dichotomy result for the complexity
  of deciding the existence of repairs of a certain size with respect to single, self-join-free denial constraints. \ \ \ . }}

 \item \red{From the repair connection we
obtain that, for consistency based-diagnosis with specifications given by positive implications with disjunctive consequents,
the problems of
computing minimum-cardinality diagnoses and computing minimum-cardinality diagnoses that contain a given
atom are both $\nit{FP}^{\nit{NP(log} (n))}$-hard in the size of their underlying Herbrand structure.} \vspace{-4mm}
\end{enumerate}
4. \ We define notions of preferred causes; in particular one based on prioritized repairs \cite{chomicki12}. We also propose an approach to causality based
  on interventions that are repair actions that replace attribute values by null values.



\vspace{2mm}
\noindent The paper is structured as follows. Section \ref{sec:prel} introduces technical preliminaries for relational databases, causality in databases, database repairs and consistent query answering, consistency-based
 diagnosis, and relevant complexity classes. Section \ref{sec:causfrepair}  characterizes actual causes and responsibilities in terms of database repairs.
Section \ref{sec:repairfcauses} characterizes repairs and consistent query answers in terms of causes and contingency sets for queries that are unions of conjunctive queries, and presents an algorithm
for computing both of the latter. \red{Section \ref{sec:MBDtoRep} formulates causality and repair problems as consistency-based diagnosis problems.} Section \ref{sec:MBDcomx} shows complexity and algorithmic results; in particular a fixed-parameter tractability result for causes' responsibilities, and also about
consistency based-diagnosis.
Section \ref{sec:pref-cause} deals with preferred causes.
 \red{Section \ref{sec:disc} discusses several relevant issues, connections and open problems around causality in databases. It also draws some final conclusions.}
 \ignore{Proofs of results without an implicit proof in this paper can be found in
\cite{icdtExt}.} We provide proofs for all the results except for those that are rather straightforward. This is an extended version of \cite{icdt15}. It contains proofs,
many improvements in the presentation, and also new developments and results, mainly in Sections \ref{sec:complMBD} and  \ref{sec:pref-cause}.

\section{ \ Preliminaries}\label{sec:prel}

We consider relational database schemas of the form $\mathcal{S} = (U,\mc{P})$,  where $U$ is the possibly infinite
database domain of {\em constants} and $\mc{P}$ is a finite set of {\em database predicates}\footnote{As opposed to built-in predicates (e.g. $\neq$) that we assume
do not appear, unless explicitly stated otherwise.} of fixed arities. A database instance $D$
compatible with $\mathcal{S}$ can be seen as a finite set of ground atomic formulas (in databases aka. atoms or tuples), of the form $P(c_1, ..., c_n)$, where $P \in \mc{P}$ has arity $n$, and $c_1, \ldots , c_n \in U$.

A {\em conjunctive query}  (CQ) is a formula $\mc{Q}(\bar{x})$ of \red{the first-order (FO) logic language, $\mc{L}(\mc{S})$, associated to $\mc{S}$} of the form \ $\exists \bar{y}(P_1(\bar{s}_1) \wedge \cdots \wedge P_m(\bar{s}_m))$,
where the $P_i(\bar{s}_i)$ are atomic formulas, i.e. $P_i \in \mc{P}$, and the $\bar{s}_i$ are sequences of terms, i.e. variables or constants\red{.\footnote{\red{In this work, we will
assume, unless otherwise explicitly said, that CQs may contain inequality atoms (equality atoms are not an issue, because they can always be eliminated).}}} The $\bar{x}$ in  $\mc{Q}(\bar{x})$ shows
all the free variables in the formula, i.e. those not appearing in $\bar{y}$. If $\bar{x}$ is non-empty, the query is {\em open}. If $\bar{x}$ is empty, the query is {\em boolean} (a BCQ), i.e. the query is a sentence, in which case, it is true or false
in a database, denoted by $D \models \mc{Q}$ and $D \not\models \mc{Q}$, respectively. A sequence $\bar{c}$ of constants is an answer to an open query $\mc{Q}(\bar{x})$ if $D \models \mc{Q}[\bar{c}]$, i.e.
the query becomes true in $D$ when the free variables are replaced by the corresponding constants in $\bar{c}$.

An {\em integrity constraint} is a sentence of language $\mc{L}(\mc{S})$, and then, may be true or false
in an instance for schema $\mc{S}$. Given a set $\IC$ of integrity constraints for schema $\mc{S}$, a database instance $D$ is {\em consistent} with $\mc{S}$ if $D \models \IC$; otherwise it is said to be {\em inconsistent}.
In this work we assume that sets of integrity constraints are always finite and logically consistent.

A particular class of  integrity constraints  is formed by {\em denial constraints} (DCs), which are sentences  $\kappa$ of the form:
 $\forall \bar{s} \neg  (A_1(\bar{s}_1)  \wedge \cdots  \wedge A_n(\bar{s}_n)$, where $\bar{s}= \bigcup \bar{s}_i$ and each $A_i(\bar{s}_i)$ is a database atom, i.e. predicate $A_i \in \mc{P}\!$.  So as with conjunctive queries, the atoms may contain
 constants.
Denial constraints  are  exactly the negations of BCQs. Sometimes we use the common representation of DCs as ``negative rules" of the form: \ $\leftarrow \ A_1(\bar{s}_1), \ldots, A_n(\bar{s}_n)$. \red{We will also consider {\em functional dependencies} (FDs) as DCs. They are represented by negative rules of the form: \ $\leftarrow \ A(\bar{x}_1,\bar{x}_2,y), A(\bar{x}_1,\bar{x}_3,z), y \neq z$, saying that the last attribute of relation $A$ functionally depends upon the attributes holding variables $\bar{x}_1$. They do not contain constants, and correspond to BCQs with inequality.}

\subsection{ \ Causality and responsibility} \label{causeandres}   Assume that the database instance is split in two, i.e. $D=D^n \cup D^x$,  where $D^n$ and $D^x$ denote the disjoint sets of {\em endogenous} and {\em exogenous} tuples,
 respectively.

\red{ Actual causes and contingent tuples are usually restricted to be among a pre-specified
set of
endogenous tuples, which are admissible, possible candidates for causes, as opposed
to the
exogenous tuples.   Actually,  the latter provide the context or the background
for the problem, and are considered as external factors that are not of interest to the
current problem statement or beyond our control.  Since no {\em intervention} (or update, in database parlance) is conceivable on
exogenous tuples, they can not be included in any contingency set or be an actual cause. They are assumed to be included in all conceivable hypothetical states of a
database.}

\red{The endogenous/exogenous partition is application-dependent and captures predetermined factors, such as users' preferences that may affect QA-causal analysis. For example,
certain tuples or full tables might be identified as irrelevant (or exogenous) in relation to
a particular query at hand, or decided to be exogenous or endogenous
a priori, independently from the query.}

A tuple $t \in D^n$ is called a
{\em counterfactual cause} for  a BCQ $\mc{Q}$,  if $D\models \mc{Q}$ and $D\smallsetminus \{t\}  \not \models \mc{Q}$.  A tuple $t \in D^n$ is an {\em actual cause} for  $\mc{Q}$
if there  exists $\Gamma \subseteq D^n$, called a {\em contingency set},  such that $t$ is a counterfactual cause for $\mc{Q}$ in $D\smallsetminus \Gamma$ \ \cite{Meliou2010a}.

We will concentrate mostly on CQs. However, the definitions of actual cause and contingency set  can be applied without a change
to {\em monotone queries} in general \cite{Meliou2010a}, in particular to {\em unions of BCQs} (UBCQs), with or without built-ins.

 The {\em responsibility} of an actual cause $t$ for $\mc{Q}$, denoted by $\rho_{_{\!D\!}}(t)$,  is the numerical value $\frac{1}{|\Gamma| + 1}$, where $|\Gamma|$ is the
size of the smallest contingency set for $t$. We can extend responsibility to all the other tuples in $D^n$ by setting their value to $0$. Those tuples are not actual causes for $\mc{Q}$.

\ignore{
In \cite{Meliou2010a}, causality for non-query answers is defined on basis of sets of {\em potentially missing tuples} that account for the missing answer. Causality for
non-answers becomes a variation of causality for answers. In this work we do not consider non-query answers.}

\begin{example}\label{ex:cfex1}
Consider $D = D^n = \{R(a_4,a_3), R(a_2,a_1), R(a_3,a_3), S(a_4),$ $S(a_2),$ $S(a_3)\}$, and the query $\mc{Q} :\exists x \exists y ( S(x) \land R(x, y) \land S(y))$. It holds: $D \models \mc{Q}$.

 Tuple $S(a_3)$ is a counterfactual cause for $\mc{Q}$. If $S(a_3)$ is removed from $D$,  $\mc{Q}$ is not true anymore. Therefore, the responsibility of $S(a_3)$ is 1. Besides, $R(a_4,a_3)$ is an actual cause for $\mc{Q}$ with contingency set
$\{ R(a_3,a_3)\}$. If $R(a_3,a_3)$ is removed from $D$, $\mc{Q}$ is still true, but further removing $R(a_4,a_3)$ makes $\mc{Q}$ false. The responsibility of $R(a_4,a_3)$ is $\frac{1}{2}$, because its smallest contingency sets have size $1$. Likewise,  $R(a_3,a_3)$ and $S(a_4)$ are actual causes for $\mc{Q}$ with responsibility  $\frac{1}{2}$.

For the  same $\mc{Q}$, but with
$D=\{S(a_3),S(a_4),R(a_4,a_3) \}$, and the
partition $D^n=\{S(a_4),S(a_3)\}$ and $D^x=\{ R(a_4,a_3)\}$, it turns out that both $S(a_3)$
and $S(a_4)$ are counterfactual
causes for $\mc{Q}$.\boxtheorem
\end{example}
\red{\begin{remark} \ In the rest of this paper, we will assume in the context of causality that database instances $D$ are
partitioned as $D=D^n \cup D^x$, into a subset of endogenous and a set of exogenous tuples, respectively. We will denote with $\nit{Causes}(D,\mc{Q})$ the set of actual causes for the BCQ  $\mc{Q}$ (being true) from instance $D$.
\end{remark}}

\subsection{ \ Database repairs}\label{sec:reps}
 Given a set $\IC$ of integrity constraints,  a {\em subset repair} (simply, S-repair) of a possibly inconsistent instance $D$ for schema $\mc{S}$  is an instance $D'$ for
$\mc{S}$ that satisfies $\IC$ and makes $\Delta(D,D')=(D \smallsetminus D') \cup( D' \smallsetminus D)$ minimal under set inclusion.\footnote{\red{In general, in the context of repairs,
partitions on instances are not considered. However, in Section \ref{sec:endo} we will bring them into the repair scene.}}
$\nit{Srep}(D,\IC)$ denotes the set of S-repairs of $D$ with respect to\ $\IC$ \cite{pods99}.
Similarly, $D'$ is a  {\em cardinality repair} (simply C-repair) of $D$ if $D'$ satisfies $\IC$ and minimizes $|\Delta(D,D')|$.
$\nit{Crep}(D,\IC)$ denotes the class of C-repairs of $D$ with respect to\ $\IC$.
C-repairs are always S-repairs.
For DCs, S-repairs and C-repairs are obtained from the original instance by deleting an S-minimal, resp. C-minimal, set of tuples. \red{In other words, S- and C-repairs under DCs
become maximal (under set inclusion), resp. maximum (in cardinality), consistent subsets of the given instance.}

\red{In more general terms, we say that a set is S-minimal in a class of sets $\mc{C}$ if it is minimal under set inclusion in $\mc{C}$.
Similarly, a set is C-minimal (or minimum) if it is minimal in cardinality within $\mc{C}$. S-minimality and C-minimality are defined similarly.}

\begin{example}\label{ex:new} \red{(ex. \ref{ex:cfex1} cont.) Consider the denial constraint \ $\kappa\!: \ \leftarrow \ S(x), R(x, y), S(y)$, whose body corresponds to the CQ in Example \ref{ex:cfex1},
and is violated by the given instance $D$. }

\red{Here, $\nit{Srep}(D, \kappa)$ $=$ $\{D_1,$ $D_2,$ $D_3\}$ with
$D_1= \{R(a_4,a_3),$ $ R(a_2,a_1),$ $ R(a_3,a_3),$ $  S(a_4),$ $ S(a_2)\}$, \ $D_2 = \{ R(a_2,a_1),$ $ S(a_4),$
$S(a_2),$ $S(a_3)\}$, \ $D_3 = \{R(a_4,a_3),$ $ R(a_2,a_1),$ $ S(a_2),$ $ S(a_3)\}$. The only C-repair is $D_1$, i.e.
 $\nit{Crep}(D, \kappa)=\{D_1\}$.} \boxtheorem
\end{example}

More generally, different {\em repair semantics} may be considered to restore consistency with respect to general integrity constraints. They depend on the kind of allowed updates on the database (i.e.
tuple insertions/deletions, changes of attribute values), and the minimality
conditions on repairs, e.g. subset-minimality, \linebreak cardinality-minimality, etc.

Given $D$ and $\IC$, a {\em repair semantics}, \red{{\large \sf S}}, defines a class \red{$\nit{Rep}^{\!\mbox{\large \sf S}\!}(D,\IC)$} of \red{{\large \sf S}-repairs}, which are
the intended repairs  \cite[Sec. 2.5]{2011Bertossi}. All the elements of $\nit{Rep}^{\!\mbox{\large \sf S}\!}(D,\IC)$ are instances over the same schema of $D$, and consistent with respect to $\IC$. If $D$ is already consistent,
$\nit{Rep}^{\!\mbox{\large \sf S}\!}(D,\IC)$ contains $D$ as its only member.

Given a repair semantics, {\large \sf S}, \ $\bar{c}$ is a {\large \sf S}-{\em consistent answer} to an open query $\mc{Q}(\bar{x})$ if $D' \models \mc{Q}[\bar{c}]$ for every $D' \in \nit{Rep}^{\!\mbox{\large \sf S}\!}(D,\IC)$.  A BCQ is {\large \sf S}-{\em consistently true} if it is true in every $D' \in \nit{Rep}^{\!\mbox{\large \sf S}\!}(D,\IC)$. In particular, if $\bar{c}$ is a consistent answer to $\mc{Q}(\bar{x})$ with respect to S-repairs, we say
 it is an {\em S-consistent answer}. Similarly for {\em C-consistent answers}.\ignore{, \red{is denoted by $D \models_S \mc{Q}[\bar{c}]$,
resp. $D \models_C \mc{Q}[\bar{c}]$}.} Consistent query answering for DCs under S-repairs was investigated in detail \cite{Chomicki05}.  C-repairs and consistent query answering under them were investigated in detail in \cite{icdt07}.
(Cf. \cite{2011Bertossi} for more references.)

\subsection{ \ Consistency-based diagnosis}\label{sec:mbd}
Consistency-based diagnosis, a form of model-based diagnosis \cite[Sec. 10.4]{struss}, considers problems $\mc{M}=(\nit{SD}, \nit{COMPS}, $ $ \nit{OBS})$, where $\nit{SD}$ is
the description in logic of the intended properties of a system under the {\em explicit} assumption that all the {\em components} in $\nit{COMPS}$  are  working normally.
$\nit{OBS}$ is a FO sentence that represents the  observations.
If the system does not behave as expected (as shown by the observations),
then the logical theory obtained from $\nit{SD} \cup \nit{OBS}$ plus the explicit assumption, say $\bigwedge_{c \in \nit{COMPS}} \neg \nit{Ab}(c)$, that the components are indeed behaving normally, becomes inconsistent.
$\nit{Ab}$ is an {\em abnormality} predicate.\footnote{Here, and as usual, the atom $\nit{Ab}(c)$ expresses that component $c$ is (behaving) abnormal(ly).}

The inconsistency is captured via the {\em minimal conflict sets}, i.e. those minimal subsets $\nit{COMPS'}$ of $\nit{COMPS}$, such that $\nit{SD} \cup \nit{OBS} \cup \{\bigwedge_{c \in \nit{COMPS'}}
\neg \nit{Ab}(c)\}$ is  inconsistent. As expected, different notions of minimality can be used at this point. \ignore{It is common to use the distinguished predicate $\nit{Ab}(\cdot)$ for denoting {\em  abnormal}
(or abnormality). So, $\nit{Ab}(c)$ says that component $c$ is abnormal.}

 A {\em  minimal diagnosis} for $\mc{M}$ is a minimal subset $\Delta$ of $\nit{COMPS}$, such that $ \nit{SD} \cup \nit{OBS} \cup \{\neg \nit{Ab}(c)~|~c \in  \nit{COMPS} \smallsetminus \Delta \} \cup \{\nit{Ab}(c)~|~c \in \Delta\}$ is consistent. That is, consistency is restored by flipping the normality assumption to abnormality for a minimal set of components, and  those are the ones considered to be (jointly) faulty. The notion of minimality commonly used is S-minimality, i.e. a diagnosis that does  not have a proper subset that is a diagnosis. We will use this kind of minimality in relation to diagnosis. Diagnosis can be obtained from conflict sets \cite{Reiter87}.

\ignore{Diagnostic  reasoning  is  non-monotonic  in  the sense  that  a  diagnosis may not survive after the addition of  new
observations \cite{Reiter87}.  }

\begin{example} \red{Consider a simple logical gate $\nit{Or}$, denoted with $o$ (the only system component in this case), that receives two digits, $x,y$, as inputs and outputs a digit
$\nit{val}(x,y)$. }

\red{This simple system can be specified in terms
of {\em normal behavior} by the logical formula
 \ $\sigma\!: \ \ \neg \nit{Ab}(o) \ \ \longrightarrow \ \ (\nit{val}(x,y) = 0 \ \longleftrightarrow \ x = y = 0)),$
saying that, when the gate is not abnormal, the output is $0$ iff the inputs are both $0$.}

\red{The logical theory \ $\{\sigma, \ \nit{val}(0,1) = 0\}$ \ is logically consistent (it can be made true) despite the unexpected observation (namely, output $0$ with inputs $0, 1$).
This is because the system's model allows for abnormal behaviors.
However, this theory together with the extra assumption $\neg \nit{Ab}(o)$, i.e. that the gate is normal, form the  theory \ $\{\sigma, \ \nit{val}(0,1) = 0, \ \neg \nit{Ab}(o)\}$
 that is inconsistent in the sense that it can not be made true
(in technical terms, it has not models).} \boxtheorem
\end{example}

\ignore{\vspace{-3mm}
\paragraph{\bf Abductive diagnosis.}  In abductive diagnosis, we have a system description $\nit{SD}$ in a certain logical formalism, very commonly
in the form of a (extended) Datalog program \cite{EiterGL97}; a set $\nit{ABD}$ of atoms, called {\em abductible atoms} (or simply, {\em abductibles}), e.g. abnormality atoms of the form $\nit{Ab}(\bar{c})$); and an observation $\nit{OBS}$ as in
consistency-based diagnosis. The abduction problem is about computing a minimal (under certain minimality criterion) $\Delta \subseteq \nit{ABD}$ such that $\nit{SD} \cup \Delta \models \nit{OBS}$. For relationships
and comparisons between consistency-based and abductive diagnosis, see \cite{Console91}.  }

\subsection{ \ Complexity classes}
We  recall some
complexity classes \cite{papa} used in this paper. $\nit{FP}$ is the class of functional problems
associated to decision problem in the class \nit{PTIME}, i.e. that are solvable
in polynomial time. $P^\nit{NP}$ (or $\Delta^P_2$) is the class of decision problems solvable in
polynomial time by a machine that makes calls to an $\nit{NP}$ oracle. For $\nit{P}^{\nit{NP(log} (n))}$
 the number of calls is logarithmic. It is not known if
$\nit{P}^{\nit{NP(log} (n))}$  is strictly contained in $\nit{P}^{\nit{NP}}\!\!$. \ $\nit{FP}^{\nit{NP(log} (n))}$ is
similarly defined.

\section{ \ Actual Causes From Database Repairs} \label{sec:causfrepair}

\red{In this section we characterize actual causes for a BCQ $\mc{Q}$ being true in a database instance $D$ in terms of the repairs  of $D$ with respect
to a denial constraint whose violation view is $\mc{Q}$, i.e. the latter asks if the constraint is violated. In essence, the actual causes will become
the tuples outside an S-repair. The complement of the latter contains the cause plus a contingency set for the cause. In order to
capture responsibility, C-repairs are considered.}

Let $D$ be an instance for schema $\mathcal{S}$, and
$\mc{Q}\!: \exists \bar{x}(P_1(\bar{x}_1) \wedge \cdots \wedge P_m(\bar{x}_m))$  a BCQ.
$\mc{Q}$ may be unexpectedly true, i.e.  $D \models \mc{Q}$.
Now, $\neg \mc{Q}$ is logically equivalent to the DC
$\kappa(\mc{Q})\!: \forall \bar{x} \neg (P_1(\bar{x}_1) \wedge \cdots \wedge P_m(\bar{x}_m))$.
 The requirement
that $\neg \mc{Q}$ holds can be captured by imposing $\kappa(\mc{Q})$ on $D$.
Due to $D \models \mc{Q}$, it holds $D \not \models \kappa(\mc{Q})$. So, $D$ is inconsistent with respect to\ $\kappa(\mc{Q})$, and could be repaired.

Repairs for (violations of) DCs are obtained
by tuple deletions.
Intuitively,  a tuple that participates in a violation of $\kappa(\mc{Q})$ in $D$ is an actual cause for $\mc{Q}$. S-minimal sets of tuples like this
are expected to correspond to S-repairs for $D$ with respect to $\kappa(\mc{Q})$.

More precisely,
given an instance $D$, a BCQ $\mc{Q}$, and a tuple $t \in D^n$, we consider:
\begin{itemize}
\item The class containing the sets of differences between $D$ and those S-repairs that do not contain $t$, and are obtained
by removing a subset of $D^n$:
\begin{eqnarray}
\nit{Diff}^s(D,\kappa(\mc{Q}), t)&=&\{ D \smallsetminus D'~|~ D' \in \nit{Srep}(D,\kappa(\mc{Q})), \nonumber\\ && \hspace*{2.5cm}t \in (D\smallsetminus D') \subseteq D^n\}.    \label{eq:df}
\end{eqnarray}
\item The class containing the sets of differences between $D$ and those C-repairs that do not contain $t$, and are obtained
by removing a subset of $D^n$:
\begin{eqnarray}
\nit{Diff}^c(D,\kappa(\mc{Q}), t)&=&\{ D \smallsetminus D'~|~ D' \in \nit{Crep}(D,\kappa(\mc{Q})),\nonumber\\ && \hspace*{2.5cm}t \in (D\smallsetminus D') \subseteq D^n\}.   \label{eq:dfc}
\end{eqnarray}
\end{itemize}

It holds $\nit{Diff}^c(D,\kappa(\mc{Q}), t) \subseteq \nit{Diff}^s(D,\kappa(\mc{Q}), t)$.

Now, any $\Lambda \in \nit{Diff}^s(D,\kappa(\mc{Q}), t)$ can be written as  $\Lambda=\Lambda' \cup \{t\}$.
From the S-minimality of S-repairs,  it follows that  $D \smallsetminus (\Lambda' \cup \{t\}) \models \kappa(\mc{Q})$, but $D \smallsetminus \Lambda' \models \neg \kappa(\mc{Q})$. That is,
$D \smallsetminus (\Lambda' \cup \{t\}) \not \models \mc{Q}$, but $D \smallsetminus \Lambda' \models \mc{Q}$. As a consequence,
$t$ is an actual cause for $\mc{Q}$ with contingency set $\Lambda'$. We have obtained the following result.\ignore{\footnote{For database instance $D$, we will usually
denote its subsets with $D'$, $\Lambda$, $\Lambda'$, $\Delta$, $\Gamma$, etc.} }

\begin{proposition}\label{pro:c&r} \em
Given an instance $D$ and a BCQ  $\mc{Q}$, $t \in D^n$ is an actual cause for $\mc{Q}$ \ iff \
$\nit{Diff}^s(D,\kappa(\mc{Q}), t) \not = \emptyset$. \red{Furthermore, if $D \smallsetminus D' \in \nit{Diff}^s(D,\kappa(\mc{Q}), t)$, then $D \smallsetminus (D' \cup \{t\})$ is
a minimal contingency set for $t$.} \boxtheorem
\end{proposition}
\begin{proposition}\label{pro:r&r} \em
Given  an instance $D$, a BCQ  $\mc{Q}$, and $t \in D^n$:
\begin{enumerate}[(a)]
\item If $\nit{Diff}^s(D,\kappa(\mc{Q}),  t) = \emptyset$, then $\rho_{_D}(t)=0$.

\item Otherwise, $\rho_{_D}(t)=\frac{1}{|\Lambda|}$, where $\Lambda \in \nit{Diff}^s(D,\kappa(\mc{Q}), t)$
and there is no $\Lambda' \in \nit{Diff}^s(D,\kappa(\mc{Q}), t)$ such that $|\Lambda'| < |\Lambda|$. \boxtheorem
\end{enumerate}
\end{proposition}
\begin{corollary}\label{cor:card} \em
Given  an instance $D$ and a BCQ  $\mc{Q}$: \  $t \in D^n$ is a most responsible actual cause for $\mc{Q}$ \ iff \
$\nit{Diff}^c(D, \kappa(\mc{Q}), t) \not = \emptyset $.\boxtheorem
\end{corollary}
\begin{example}\label{ex:CausASrepex1} (ex. \ref{ex:cfex1} and \ref{ex:new} cont.) Consider the same instance $D$ and query $\mc{Q}$. \red{The associated DC  is  \ $\kappa(\mc{Q})\!: \ \leftarrow S(x),R(x, y),S(y)$ that we considered in Example \ref{ex:new}, where we obtained \
$\nit{Srep}(D, \kappa(\mc{Q}))$ $=$ $\{D_1,$ $D_2,$ $D_3\}$ and $\nit{Crep}(D, \kappa(\mc{Q}))=\{D_1\}$.}

For tuple $R(a_4,a_3)$, $\nit{Diff}^s(D,\kappa(\mc{Q}), R(a_4,a_3))=\{D \smallsetminus D_2\} = \{ \{ R(a_4,a_3),$ $R(a_3 $ $,a_3)\} \}$, which, by Propositions  \ref{pro:c&r} and \ref{pro:r&r},
confirms that $R(a_4,a_3)$ is an actual cause,  with responsibility $\frac{1}{2}$. \red{The complement
 of  $D\smallsetminus D_2$  contains the actual cause $R(a_3,a_3)$ plus a contingency set of it, namely that formed by tuple $R(a_3,a_3)$, which has to be deleted
 together with the actual cause $R(a_4,a_3)$ to restore consistency (cf. Example \ref{ex:new}).}

 For tuple $S(a_3)$,  $\nit{Diff}^s(D,\kappa(\mc{Q}), $ $ S(a_3)) = \{D \smallsetminus D_1\}$ $=\{ S(a_3) \}$. So, $S(a_3)$
is an actual cause  with responsibility 1.

Similarly, $R(a_3,a_3)$ is an actual cause with responsibility $\frac{1}{2}$, because  $\nit{Diff}^s(D,$ $\kappa(\mc{Q}),$
 $R(a_3,a_3)) = \{D\smallsetminus D_2, \ D \smallsetminus D_3\}$ $=\{ \{R(a_4,$ $a_3),$ $R(a_3,a_3)\},$ $ \{R(a_3,a_3),$ $ S(a_4)\} \}$.

It holds  $\nit{Diff}^s(D,\kappa(\mc{Q}),S(a_2)) = \nit{Diff}^s(D,\kappa(\mc{Q}),R(a_2 , a_1)) = \emptyset$, because all repairs contain $S(a_2)$, $R(a_2 , a_1)$.
This means they do not participate in the violation of $\kappa(\mc{Q})$ or contribute to make $\mc{Q}$ true.
 So, they are not actual causes for $\mc{Q}$, confirming the result in Example \ref{ex:cfex1}.

$\nit{Diff}^c(D,\kappa(\mc{Q}), S(a_3))=\{S(a_3)\}$. From Corollary \ref{cor:card},
$S(a_3)$  is the most responsible cause. \boxtheorem
\end{example}

\begin{remark} \label{rem:ucq}
\red{The results in this section can be easily extended to unions of BCQs. This can be done by associating a DC to each disjunct of the query, and considering the corresponding problems for database repairs with respect to several DCs (cf. Section \ref{sec:disjcauses}).}
\boxtheorem
\end{remark}

\section{ \ Database Repairs From Actual Causes} \label{sec:repairfcauses}

\red{In this section we characterize repairs for inconsistent databases with respect to\ {\em a
set of} DCs in terms of actual causes with their contingency sets. The reduction of repair-related computations to cause-related computations is particularly relevant, because
we can take advantage of known complexity results for repairs to obtain new lower-bound complexity results for causality. }

\red{Causality has been investigated so far mainly for single conjunctive queries. However,
database repairs appear in the context of {\em sets} of constraints. We concentrate on sets of DCs, which requires extending the analysis of causality to {\em unions
of conjunctive queries}.}

\red{More concretely, in this section we characterize repairs of a database instance $D$ with respect to a set $\Sigma$ of DCs in terms of the actual causes (with their contingency sets)
for
the union of the conjunctive queries naturally associated to the (bodies of the) DCs.  In essence, an  S-repair $D'$ is a maximal subset of $D$ that does not contain any actual cause $t$, and
the tuples other than $t$ and outside $D'$ form a contingency set for $t$. As expected, C-repairs require the use of most responsible tuples.}

 Consider an instance $D$  for  schema $\mathcal{S}$, and a set of
DCs $\Sigma$ on $\mc{S}$.  For each $\kappa \in \Sigma$, say \
 $\kappa\!: \ \leftarrow A_1(\bar{x}_1),\ldots,A_n(\bar{x}_n)$, consider its associated  {\em violation view} defined by a BCQ, namely \
 $V^{\!\kappa}\!\!: \
\exists\bar{x}(A_1(\bar{x}_1)\wedge \cdots \wedge A_n(\bar{x}_n))$. The answer {\em yes} to $V^{\!\kappa}$ shows that $\kappa$ is violated (i.e. not satisfied)
by $D$.

Next, consider the query that is
the union of the individual violation views:  \
$V^{\!\Sigma}:= \bigvee_{\kappa \in  \Sigma} V^\kappa $, a {\em union of} BCQs (UBCQs).
Clearly, $D$
violates (is inconsistent with respect to)  $\Sigma$ iff $D \models V^{\!\Sigma}\!\!$.

 It is easy to verify that $D$, with $D^x = \emptyset$, is consistent with respect to\ $\Sigma$ \ iff \
$\nit{Causes}(D,$ $V^{\!\Sigma}) = \emptyset$, i.e. there are no actual causes for $V^{\!\Sigma}$
when all tuples are endogenous.

Now, let us collect all
{\em S-minimal contingency sets} associated with an actual cause $t$ for
$V^{\!\Sigma}$:
\begin{definition}\label{def:cont} For an instance $D$ and  a set $\Sigma$ of DCs:
\begin{eqnarray}
\nit{Cont}(D,V^{\!\Sigma},t) &:=& \{  \Gamma\subseteq D^n~|~D\smallsetminus \Gamma
\models V^{\!\Sigma}, ~D\smallsetminus (\Gamma \cup \{t\}) \not \models V^{\!\Sigma}, \label{eq:ct}
\\
&& ~~~~~~~~~~~~~~  \mbox{ and } \forall \Gamma'\subsetneqq \Gamma, \ D \smallsetminus (\Gamma' \cup \{t\})
\models V^{\!\Sigma} \}. \hspace{10mm} \Box \nonumber
\end{eqnarray}
\end{definition}

Notice that for  $\Gamma \in \nit{Cont}(D,V^{\!\Sigma},t)$, it holds $t \notin \Gamma$.
When $D^x = \emptyset$, if $t \in \nit{Causes}(D,V^{\!\Sigma})$ and $\Gamma \in \nit{Cont}(D,V^{\!\Sigma},t)$,  from  the
definition of actual cause and the S-minimality of
$\Gamma$, it holds that $\Gamma''= \Gamma \cup \{t\}$ is an S-minimal subset of $D$
with $D \smallsetminus \Gamma'' \not \models V^{\!\Sigma}$. So,  $D
\smallsetminus \Gamma''$
is an S-repair for $D$.  Then, the following holds.
\begin{proposition}\label{pro:sr&cp} \em For an instance $D$, with $D^x = \emptyset$, and a set DCs
$\Sigma$: \
$D' \subseteq D$ is an S-repair for $D$ with respect to\ $\Sigma$ iff,  for every $t
\in D \smallsetminus D'$:
$t \in \nit{Causes}(D, V^{\!\Sigma})$ and $D \smallsetminus (D' \cup
\{t\}) \in \nit{Cont}(D, V^{\!\Sigma}, t)$.
\boxtheorem
\end{proposition}

To establish a connection between most responsible actual causes and
C-repairs, assume that $D^x = \emptyset$, and collect the {\em most responsible actual causes} for
$V^{\!\Sigma}$:

\begin{definition}\label{def:mrc} For an instance $D$ with $D^x = \emptyset$:
\begin{eqnarray}
\nit{MRC}(D,V^{\!\Sigma}) &:=& \{t \in D~|~ t \in
\nit{Causes}(D,V^{\!\Sigma}),  \not \exists t' \in
\nit{Causes}(D,V^{\!\Sigma}) \label{eq:mrc}\\
&& \hspace*{1.2cm}\ \mbox{ with }  \rho_{_D}(t')> \rho_{_D}(t)\}. \hspace{4.3cm} \Box \nonumber
\end{eqnarray}
\end{definition}
\begin{proposition}\label{pro:cr&mrp} \em
For instance $D$, with $D^x = \emptyset$,  and   set of DCs $\Sigma$: \ $D' \subseteq D$ is a
C-repair for $D$ with respect to\ $\Sigma$ \ iff,  for every $t \in D \smallsetminus D'$:
$t \in  \nit{MRC}(D, V^{\!\Sigma})$ and $D \smallsetminus (D' \cup \{t\}) \in
\nit{Cont}(D, V^{\!\Sigma}, t)$.
\boxtheorem
\end{proposition}

Actual causes for $ V^{\!\Sigma}$,  with their contingency
sets, account for the violation of some $\kappa \in \Sigma$. Removing those tuples from
$D$ should remove the inconsistency. From Propositions \ref{pro:sr&cp} and \ref{pro:cr&mrp} we obtain: \vspace{-1mm}
\begin{corollary}\label{col:sr&cp} \em
Given an instance $D$ and a set DCs
$\Sigma$,  the instance obtained from $D$ by
removing an actual cause, resp.  a most responsible actual cause, for $ V^{\!\Sigma}$ together with any of its S-minimal, resp. C-minimal, contingency
sets forms an S-repair, resp. a C-repair, for $D$ with respect to $\Sigma$.
\boxtheorem
\end{corollary}
\begin{example}\label{ex:rc2cp}
 Consider $D=\{P(a),P(e),Q(a,b),R(a,c)\}$ and $\Sigma=\{\kappa_1,\kappa_2\}$, with
$\kappa_1\!: \ \leftarrow P(x), Q(x,y)$ and $\kappa_2 \!: \ \leftarrow
P(x), R(x,y)$.

The violation views  are $ V^{\kappa_1}\!: \exists xy (P(x) \land Q(x,y)) $ and
 $ V^{\kappa_2} \!: \exists xy (P(x) \land R(x,y))$.
For $V^{\!\Sigma} :=  V^{\kappa_1}\lor V^{\kappa_2}$,
$D \models V^{\!\Sigma}$ and $D$ is inconsistent with respect to\ $\Sigma$.

Now assume  all tuples are endogenous. It holds  $\nit{Causes}(D,  V^{\!\Sigma})=\{P(a),$ $Q(a,b),R(a,c) \}$, and its
elements are  associated with sets of S-minimal contingency sets, as follows: \
$\nit{Cont}(D,V^{\!\Sigma}\!,Q(a,b))=\{\{ R(a,c)\}\}$, \
$\nit{Cont}(D,V^{\!\Sigma}\!,$ $R(a,c))=\{ \{Q(a,b)\}\}$, \ and
$\nit{Cont}(D,V^{\!\Sigma}\!,P(a))=\{\emptyset\}$.

From Corollary \ref{col:sr&cp},  and $\nit{Cont}(D, V^{\!\Sigma}\!,$ $R(a,c)) = \{ \{Q(a,b)\}\}$,   $D_1=D \smallsetminus (\{
R(a,c)\} \cup \{Q(a,b)\}) =\{ P(a),P(e)\}$  is an S-repair. So is
$D_2=D \smallsetminus (\{P(a) \} \cup \emptyset)=\{P(e),Q(a,b),$ $R(a,c)\}$. These are the only S-repairs.

Furthermore, $\nit{MRC}(D, V^{\!\Sigma})= \{P(a)\}$.  From
Corollary \ref{col:sr&cp}, $D_2$ is also a C-repair for $D$.
\boxtheorem
\end{example}

\red{\begin{remark} An actual cause $t$ with any of its S-minimal contingency sets determines a unique S-repair. The last example shows that, with different combinations of a cause and one of its contingency sets, we may obtain the same repair
(e.g. for the first two $\nit{Cont}$ sets).
 So, we may have more minimal contingency sets than minimal repairs. However,   we may still have exponentially many minimal
 contingency sets, so as we may have exponentially many minimal repairs of an instance with respect to DCs, as the following example shows.\footnote{Cf. \cite{Arenas03} for an example of the latter
 that uses key constraints, which are DCs with inequalities (with violation views that contain inequality).} \boxtheorem
 \end{remark}}

\begin{example}\label{ex:repsVSconts}
Consider $D = \{R(1,0), R(1,1), \ldots, R(n,0), R(n,1), S(1), S(0)\}$ and the DC $\kappa\!: \ \leftarrow R(x,y), R(x,z), S(y),S(z)$. $D$ is inconsistent with respect to $\kappa$. There are exponentially many S-repairs of $D$: \
$D' = D \smallsetminus \{S(0)\}$,
 $D'' = D \smallsetminus \{S(1)\}$, $D_1 = D \smallsetminus \{R(1,0), \ldots, R(n,0)\}$, ..., $D_{2^n} = D \smallsetminus \{R(1,1), \ldots, R(n,1)\}$.
 The C-repairs are only $D'$ and $D''$.

For the BCQ $V^\kappa$ associated to $\kappa$, $D \models V^\kappa$, and $S(1)$ and $S(0)$ are actual causes for
$V^\kappa$ (courterfactual causes with responsibility $1$). All tuples in $R$ are  actual causes, each with exponentially many S-minimal contingency sets. For example,
$R(1, 0)$ has the S-minimal contingency set $\{R(2,0), \ldots, R(n,0)\}$, among  exponentially many others (any set
built with just one element from each of the pairs $\{R(2,0),R(2,1)\}$, ...,  $\{R(n,0), R(n,1)\}$  is one). \ignore{Therefore, $R(1, 0)$ has $2^{n-1}$
contingency sets. A similar argument holds for the rest of tuples in $R$.} \boxtheorem
\end{example}

\subsection{ \ Causes for unions of conjunctive queries}\label{sec:disjcauses}

If we want to compute repairs with respect to sets of DCs from causes for UBCQs using, say Corollary \ref{col:sr&cp}, we first
need an algorithm for computing the actual causes
and their (minimal) contingency sets for  UBCQs.
These algorithms could be used as a first stage
of the  computation of S-repairs and C-repairs with respect to sets of DCs. However, these algorithms (developed in Section
\ref{sec:disjcausesCont}), are also interesting and useful {\em per se}.

The PTIME algorithm for computing actual causes in \cite{Meliou2010a} is for single conjunctive queries, but does not
compute the actual causes' contingency sets.
 Actually, doing the latter increases the complexity, because {\em deciding responsibility}\footnote{For a precise formulation, see
 Definition \ref{def:resp}.} of actual causes is $\nit{N\!P}$-hard \cite{Meliou2010a} (which would be tractable
 if we could efficiently compute all (minimal) contingency sets).\footnote{Actually, \cite{Meliou2010a} presents a PTIME algorithm
for computing responsibilities  for a restricted class of CQs.} In principle, an algorithm for responsibilities can be used to compute C-minimal contingency sets, by iterating over all candidates,
but Example \ref{ex:repsVSconts} shows that there  can be exponentially many of them.

We first concentrate on the problem of computing actual causes for UBCQs, without their contingency sets, which requires some notation.
\begin{definition} \label{def:hsStuff}  Given  $\mc{Q} =  C_1 \vee \cdots \vee C_k$, where each $C_i$ a BCQ, and an instance $D$:
\begin{enumerate}[(a)]
\item $\mathfrak{S}(D)$ is the collection of all
 S-minimal subsets of $D$ that satisfy a
disjunct  $C_i $ of $\mc{Q}$.

\item $\mf{S}^n(D)$ consists of the S-minimal subsets $\Lambda$ of  $D^n$ for which there exists a  $\Lambda' \! \in \mf{S}(D) $ with
$\Lambda \subseteq \Lambda'$ and $ \Lambda \smallsetminus \Lambda' \subseteq D^x$. \boxtheorem
\end{enumerate}
\end{definition}

$\mf{S}^n(D)$  contains all S-minimal sets of endogenous tuples that simultaneously (and possibly accompanied by exogenous tuples) make the query true.
It is easy to see that  $\mathfrak{S}(D)$ and $\mf{S}^n(D)$ can be computed in polynomial time in the size of $D$.

Now, generalizing a result for CQs in \cite{Meliou2010a}, actual causes for a UBCQs can be computed in PTIME in the size of $D$ without computing contingency
 sets. We formulate this results in terms of the corresponding {\em causality decision problem} (CDP).
\begin{proposition}\label{pro:UBCQCausesindirect} \em
Given an instance $D$, a UBCQ $\mc{Q}$, and $t \in D^n$:
\begin{enumerate}[(a)]
\item $t$ is an actual cause for $\mc{Q}$ iff  there is  $\Lambda  \in   \mf{S}^n(D)$ with $t \in \Lambda$.

\item The {\em causality decision problem} (about membership of)
\begin{equation}\label{eq:cpd}
\mc{CDP} := \{ (D,t)~|~ t \in D^n, \mbox{ and } t \in \nit{Causes}(D,\mc{Q})\}
\end{equation}
 belongs to $\nit{PTIME}$.
\end{enumerate}
\end{proposition}

\proof{(a) Assume $\mf{S}(D) =\{\Lambda_1, \ldots, \Lambda_m\}$, and there exists a $\Lambda \in\mf{S}^n(D)$ with $t \in \Lambda$. Consider a set $\Gamma \subseteq D^n$ such that, for all $\Lambda_i \in\mf{S}^n(D)$ where $\Lambda_i \not = \Lambda$,  $\Gamma \cap \Lambda_i \not = \emptyset$ and  $\Gamma \cap \Lambda =\emptyset$. With such a $\Gamma$,
     $t$ is an actual cause for $\mc{Q}$ with contingency set $\Gamma$. So, it is good enough to prove that such $\Gamma$ always exists. In fact, since all subsets of $\mf{S}^n(D)$ are S-minimal, then, for each $\Lambda_i \in\mf{S}^n(D)$ with $\Lambda_i \not = \Lambda$, $\Lambda_i \cap \Lambda = \emptyset$. Therefore, $\Gamma$ can be obtained from the set of difference between each $\Lambda_i$ and $\Lambda$.

     Now, if $t$ is an actual cause for $\mc{Q}$, then there exist an S-minimal $\Gamma \in D^n$, such that  $D   \smallsetminus (\Gamma \cup\{t\}) \not \models \mc{Q}$, but  $D  \smallsetminus \Gamma \models \mc{Q}$. This implies that there exists an S-minimal subset $\Lambda$ of $D$, such that $t \in \Lambda
 $ and $\Lambda \models \mc{Q}$. Due to the S-minimality of $\Gamma$, it is easy to see that $t$ is included in a subset of $\mf{S}^n(D)$.

 \noindent (b) This is a simple generalization of the proof of the same result for single conjunctive queries found in \cite{Meliou2010a}. \boxtheorem\\}

\begin{example}\label{ex:al2} (ex. \ref{ex:rc2cp} cont.) \ Consider the query $\mc{Q}\!: \  \exists xy (P(x)
\land Q(x,y)) \lor \exists xy(P(x) \land R(x,y))$, and assume that for $D$, \ $D^n=\{P(a), R(a,c)\}$ and $D^x=\{P(e),Q(a,b)\}$. It holds $\mf{S}(D)=\{ \{P(a),Q(a,b)\}, \{P(a),$
$R(a,c)\} \}$.  Since  $\{P(a)\} \subseteq \{P(a),$
$R(a,c)\}$, \ $\mf{S}^n(D)=\{\{P(a)\}\}$. So, $P(a)$ is the only actual cause for $ \mc{Q}$.
\boxtheorem
\end{example}

\subsection{ \ Contingency sets for unions of conjunctive queries}\label{sec:disjcausesCont}
 It is possible to develop a (naive) algorithm  that accepts as input an instance
$D$ and a UBCQ $\mc{Q}$,  and returns
$\nit{Causes}(D,\mc{Q})$; and also, for
each $t \in \nit{Causes}(D,\mc{Q})$, \ its (set of) S-minimal contingency sets $\nit{Cont}(D,\mc{Q},t)$.

The basis for the algorithm is a correspondence between the actual causes for $\mc{Q}$ with their
contingency sets and a {\em hitting-set problem}.\footnote{If $\mc{C}$ is a collection of non-empty subsets of a set $S$, a subset $S' \subseteq S$ is a {\em hitting-set} for
    $\mc{C}$  if, for every  $C \in \mc{C}$, $C \cap S' \neq \emptyset$. \ $S'$ is an S-minimal hitting-set if no proper subset of
    it is also a hitting-set. $S$ is a minimum hitting-set if it has minimum cardinality.}
More precisely, for a fixed UBCQ $\mc{Q}$,  consider the {\em hitting-set framework}
\begin{equation}\label{eq:frame}
\mf{H}^n(D) = \langle D^n, \mf{S}^n(D)\rangle,
\end{equation} with  $\mf{S}^n(D)$ as in Definition
\ref{def:hsStuff}. Different computational and decision problems are based on $\mf{H}^n(D)$, and we will confront some below.  Notice that hitting-sets (HSs) are all subsets of $D^n$.

The  S-minimal hitting-sets for $\mf{H}^n(D)$ correspond to actual causes with their S-minimal contingencies for $\mc{Q}$. Most responsible causes for $\mc{Q}$ are in correspondence with hitting-sets  for $\mf{H}^n(D)$.
This is formalized as follows:
\begin{proposition}\label{pro:UBCQCauses} \em
For an instance $D$, a UBCQ $\mc{Q}$, and $t \in D^n$:
\begin{enumerate}[(a)]
\item $t$ is an actual cause for $\mc{Q}$ with S-minimal contingency set $\Gamma$ iff  $\Gamma \cup \{t\}$ is an S-minimal hitting-set for  $\mf{H}^n(D)$.

\item $t$ is a most responsible actual cause for $\mc{Q}$ with C-minimal contingency set $\Gamma$ iff  $\Gamma \cup \{t\}$ is a minimum hitting-set  for $\mf{H}^n(D)$.\boxtheorem
\end{enumerate}
\end{proposition}

The proof is similar to that of part (a) of Proposition \ref{pro:UBCQCausesindirect}.
\begin{example}\label{ex:al1} (ex. \ref{ex:rc2cp} and \ref{ex:al2} cont.) \  $D$ and $\mc{Q}$ are as before, but now all tuples are endogenous.
 Here,
$\mf{S}(D)= \mf{S}^n(D) = \{ \{P(a),Q(a,b)\}, \{P(a),$
$R(a,c)\} \}$. $\mf{H}^n(D)$ has two S-minimal hitting-sets:
$H_1=\{P(a)\}$ and $H_2=\{Q(a,b), R(a,c)\}$. Each of them implicitly contains an
actual cause (any of its elements)  with an S-minimal  contingency set (what's left after removing the actual cause). $H_1$ is also the C-minimal hitting-set, and contains the most responsible actual cause,
$P(a)$.
\boxtheorem
\end{example}

\begin{remark} \em
\label{rem:hs}\em For $\mf{H}^n(D)=\langle D^n, \mf{S}^n(D)\rangle$, $\mf{S}^n(D)$ can be computed in PTIME \red{in data complexity}, and its elements are bounded  in size by $|\mc{Q}|$,  which is the
maximum number of atoms in one of $\mc{Q}$'s disjuncts. This is a special kind of hitting-set problems. For example, deciding if there is a hitting-set of size at most $k$ as been called the $d$-{\em hitting-set problem} \cite{NiedermeierR03}, and $d$ is the bound on the size of the sets in the set class. In our case, $d$ would be $|\mc{Q}|$. \boxtheorem
\end{remark}

\subsection{ \ Causality, repairs, and consistent answers} \label{sec:cqa}

Corollary \ref{col:sr&cp} and  Proposition \ref{pro:UBCQCauses} can be used to compute repairs. If the classes of S- and C-minimal hitting-sets for $\mf{H}^n(D)$ (with $D^n = D$) are available, computing S- and C-repairs will be in PTIME
in the sizes of those classes. However, it is well known that computing minimal hitting-sets is a complex problem. Actually, as Example \ref{ex:repsVSconts} implicitly shows, we can have exponentially many of them in  $|D|$; so
 as exponentially many minimal repairs for $D$  with respect to a denial constraint. We can see that the complexity of contingency sets computation is in line with the complexities of
 computing hitting-sets and repairs.

As Corollary \ref{col:sr&cp} and Proposition \ref{pro:UBCQCauses} show, the  computation of causes, contingency sets, and most responsible causes via minimal/minimum hitting-set computation can be used to compute repairs and decide about repair questions. Since the
 hitting-set problems in our case are of the $d$-hitting-set kind, good algorithms and approximations for the latter (cf. Section \ref{sec:fpt}) could be used in the context of repairs.

In the rest of this section we consider an instance $D$ whose tuples are all  endogenous,  and a set $\Sigma$ of DCs.
\ For the disjunctive violation view $V^{\!\Sigma}$, the following result is obtained from Propositions \ref{pro:sr&cp} and \ref{pro:cr&mrp}, and Corollary \ref{col:sr&cp}. \vspace{-1mm}
\begin{corollary}\label{col:consinf} \em
For an instance $D$, with $D^x = \emptyset$, and a set  $\Sigma$ of DCs, it holds:
\begin{enumerate}[(a)]
\item  For every
$t \in \nit{Causes}(D,V^{\!\Sigma})$, there is an S-repair that does not contain $t$.

\item  For every
$t \in \nit{MRC}(D, V^{\!\Sigma})$, there is a C-repair that does not contain $t$.

\item For every $D' \in \nit{Srep}(D,\Sigma)$ and $D'' \in \nit{Crep}(D,\Sigma)$,  it holds
$D \smallsetminus D' \subseteq \nit{Causes}(D,V^{\!\Sigma})$ and $D \smallsetminus D'' \subseteq \nit{MRC}(D,V^{\!\Sigma})$.
\boxtheorem
\end{enumerate}
\end{corollary}

For a
projection-free, and a possibly non-boolean CQ $\mc{Q}$,
we are interested in its consistent answers from $D$ with respect to $\Sigma$. For example, for
$\mc{Q}(x,y,z)\!: \ R(x,y) \wedge S(y,z)$, the S-consistent (C-consistent) answers would be of the form
$\langle a,b,c \rangle$, where $R(a,b)$ and $S(b,c)$ belong to
all S-repairs (C-repairs) of $D$.

From Corollary \ref{col:consinf}, $\langle a,b,c\rangle$ is an S-consistent (resp. C-consistent) answer iff $R(a,b)$ and $S(b,c)$ belong to $D$, but they are  not
actual causes (resp. most responsible actual causes) for  $ V^{\!\Sigma}$.

The following simple result and its corollary will be useful in Section \ref{sec:MBDcomx}.

\begin{proposition}\label{pro:cqa} \em
For an instance $D$, with $D^x = \emptyset$, a set $\Sigma$ of DCs, and a projection-free
CQ \ $\mc{Q}(\bar{x})\!: \  P_1(\bar{x}_1) \wedge \cdots
\wedge P_k(\bar{x}_k)$:
\begin{enumerate}[(a)]
\item $\bar{c}$ is an S-consistent answer iff, for each $i$,  $P_i(\bar{c}_i) \in (D \smallsetminus \nit{Causes}(D, V^{\!\Sigma}))$.

\item
$\bar{c}$ is a C-consistent answer iff, for each $i$,
$P_i(\bar{c}_i) \in (D
\smallsetminus \nit{MRC}(D, V^{\!\Sigma}))$.
\boxtheorem
\end{enumerate}
\end{proposition}
\begin{example}\label{ex:cqa1} (ex. \ref{ex:rc2cp} cont.) \ Consider
  $\mc{Q}(x)\!: \ P(x)$.   We had
$\nit{Causes}(D, V^{\!\Sigma})$ $=\{P(a),$ $Q(a,b),$ $R(a,c) \}$,
$\nit{MRC}(D, V^{\!\Sigma})= \{P(a)\}$. Then,  $\langle e\rangle$ is both an S- and a C-consistent answer. \boxtheorem
\end{example}

Notice that Proposition \ref{pro:cqa} can easily be
extended to conjunctions of ground atomic queries.
\begin{corollary}\label{cor:cqa&cox} \em
Given an instance $D$ and a set $\Sigma$ of DCs, the ground atomic query $\mc{Q}\!\!:  P(c)$ is C-consistently true
iff  $P(c) \in D$ and it is not a most responsible cause for $V^{\!\Sigma}$.\boxtheorem
\end{corollary}


\begin{example}\label{ex:cqa2} For  $D=\{P(a,b),R(b,c),R(a,d)\}$ and the DC $\kappa\!: \ \leftarrow P(x, y),R(y, z)$, we obtain: \
$\nit{Causes}(D,$ $V^{\kappa})=\nit{MRC}(D, V^{\kappa})=\{P(a,b),R(b,c)\}$.

From Proposition \ref{pro:cqa}, the ground atomic query
$\mc{Q}\!\!: R(a,d)$ is
both S- and C-consistently true in $D$ with respect to $\kappa$,
because, $D \smallsetminus \nit{Causes}(D, V^{\kappa}) = D \smallsetminus
\nit{MRC}(D, V^{\kappa})$ $= \{R(a,d)\}$.
\boxtheorem
\end{example}

The CQs considered in  Proposition \ref{pro:cqa} and its Corollary \ref{cor:cqa&cox} are not
 particularly interesting {\em per se}, but we will use those results to obtain new complexity results for causality later on, e.g.
Theorem \ref{the:cqa&ca&cox}.

\section{ \ \red{Causes and Repairs from Consistency-Based Diagnosis}} \label{sec:MBDtoRep}

\red{The main objective in this section is to characterize database causality computation as a diagnosis problem.\red{\footnote{\red{The other direction is beyond the scope of this work.
More importantly, logic-based diagnosis in general is a much richer scenario than that of database causality. In the former, we can have arbitrary logical specification, whereas
under data causality, we have only monotone queries at hand.}}}  This is interesting {\em per se},
and will also allow us to apply ideas and techniques from model-based diagnosis to causality. As a side result we obtain a characterization
of database repairs  in terms of diagnosis. }

Let $D$ be an instance for schema $\mathcal{S}$, and
$\mc{Q}\!: \exists \bar{x}(P_1(\bar{x}_1) \wedge \cdots \wedge P_m(\bar{x}_m))$, a  BCQ.
Assume $\mc{Q}$ is, possibly  unexpectedly, true in  $D$. So, for the associated
DC $\kappa(\mc{Q})\!: \forall \bar{x} \neg (P_1(\bar{x}_1) \wedge \cdots \wedge P_m(\bar{x}_m))$,
$D \not \models \kappa(\mc{Q})$. $\mc{Q}$ is our {\em observation}, for which we want to find explanations, using a consistency-based diagnosis approach.

For each predicate $P \in \mc{P}$, we introduce predicate $\nit{Ab}_P$, with the same arity as $P$. Intuitively, a tuple
in its extension is  {\em abnormal} for $P$.
The ``system description", $\nit{SD}$, includes, among other elements,
the original database, expressed in logical terms, and the DC as true ``under normal conditions".

More precisely, we consider the following {\em diagnosis problem},  $\mathcal{M}=(\nit{SD},D^n,$ $ \mc{Q})$, associated to $\mc{Q}$. The FO system description, $\nit{SD}$,
contains the following elements:

\begin{enumerate}[(a)]
\item  $\nit{Th}(D)$, which is
Reiter's logical reconstruction of $D$ as a FO theory \cite{Reiter82} (cf. Example \ref{ex:mbdaex5}).

\item Sentence $\kappa(\mc{Q}){\!^\nit{Ab}}$, which is $\kappa(\mc{Q})$ rewritten as follows:
\begin{eqnarray}
\kappa(\mc{Q}){\!^\nit{Ab}}\!:  \ \forall   \bar{x}\neg (P_1(\bar{x}_1)  \wedge \neg \nit{Ab}_{P_1}(\bar{x}_1)  \wedge \cdots \wedge
P_m(\bar{x}_m) \wedge \neg \nit{Ab}_{P_m}(\bar{x}_m) ). \label{eq:ext}
\end{eqnarray}
\item Formula (\ref{eq:ext}) can be refined by applying the abnormality predicate, $\nit{Ab}$,
to endogenous tuples only. For this we need to use additional auxiliary predicates $\nit{End}_{\!P}$, with the same arity of $P \in \mc{S}$, which contain the endogenous
tuples in $P$'s extension (see Example \ref{ex:mbdaex5}). \red{Accordingly, we introduce the inclusion dependencies: \ For each $P \in \mc{P}$,
$$\forall \bar{x}(\nit{Ab}_P(\bar{x}) \rightarrow \nit{End}_{\!P}(\bar{x})), \ \mbox{ and } \ \forall \bar{x}(\nit{End}_{\!P}(\bar{x}) \rightarrow P(\bar{x})).$$}
\end{enumerate}

The last entry,  $\mc{Q}$, in $\mathcal{M}$  is the ``observation", which together with \nit{SD} will produce  and inconsistent theory,
because  we make the initial and explicit assumption that all the abnormality predicates are empty (equivalently, that  all tuples are normal), i.e. we
consider, for each predicate $P$, the sentence\footnote{Notice that these can also be seen as DCs, since they can be written as $\forall \bar{x} \neg \nit{Ab}_P(\bar{x})$.}
\begin{equation} \label{eq:default}
\forall \bar{x}(\nit{Ab}_P(\bar{x}) \rightarrow \mbox{\bf false}),
\end{equation}
 where, {\bf false} is a propositional atom that is always false.

The second entry in $\mc{M}$ is $D^n$. This is the set of ``components" that we can use to try to restore consistency, in this case, by (minimally) changing the abnormality condition on tuples in $D^n$. In other words, the universal rules (\ref{eq:default})
are subject to exceptions or qualifications: some endogenous tuples may be abnormal. Each diagnosis  shows an S-minimal set of endogenous tuples that are abnormal.

\begin{example}\label{ex:mbdaex5} (ex. \ref{ex:cfex1} cont.) \ Consider the query $\mc{Q} :\exists x \exists y ( S(x) \land R(x, y) \land S(y))$, and the instance $D=\{S(a_3),$ $S(a_4),$ $R(a_4,a_3) \}$, with $D^n$ $=$ $\{S(a_4),$ $S(a_3)\}$, consider the diagnostic problem
 $\mathcal{M}=( \nit{SD},\{S(a_4),S(a_3)\},$ $ \mc{Q})$, with
$\nit{SD}$ containing the sentences in (a)-(c) below:

\begin{enumerate}[(a)]
\item Predicate completion axioms plus {\em unique names assumption}:
\begin{eqnarray}
&&\forall x y (R(x,y) \leftrightarrow x= a_4 \wedge y = a_3), \ \ \ \forall x(S(x) \leftrightarrow x = a_3 \vee x = a_4), \label{eq:one}\\
&& \forall x y (\nit{End}_R(x,y) \leftrightarrow \mbox{\bf false}), \ \ \ \forall x(\nit{End}_S(x) \leftrightarrow x = a_3 \vee x = a_4),\\
&& a_4 \neq a_3.\label{eq:tres}
 \end{eqnarray}
\item The denial constraint qualified by non-abnormality, $\kappa(\mc{Q})^{\!\nit{Ab}}$:
\begin{eqnarray*}
\forall x y \neg&&\!\!\!\!( S(x) \land  \neg   \nit{Ab}_S(x) \land  R(x, y)  \land \neg   \nit{Ab}_R(x, y)\land \ S(y)  \wedge \neg   \nit{Ab}_S(y)).
\end{eqnarray*}
In diagnosis formalizations this formula would be usually presented as:
\begin{eqnarray}
\forall x y(&&\!\!\!(\neg\nit{Ab}_S(x) \wedge \neg\nit{Ab}_R(x,y) \wedge \neg \nit{Ab}_S(y)) \ \longrightarrow 
\neg (S(x) \land  R(x, y) \land  \ S(y))).\nonumber
\end{eqnarray}
That is, under the normality assumption, the ``system" behaves as intended; in this case, there are no \ignore{(endogenous)} violations of the denial constraint.
This main formula in the diagnosis specification can also be written as a  disjunctive positive rule:
\begin{eqnarray}
&\hspace*{-7mm}\!\!\!\red{\forall x y(S(x) \land  R(x, y) \land  S(y)  \ \longrightarrow \ \nit{Ab}_S(x) \vee \nit{Ab}_R(x,y) \vee \nit{Ab}_S(y)).}\label{eq:disj}
\end{eqnarray}
\item Abnormality/endogenousity predicates are in correspondence to the database schema, and only endogenous tuples can be abnormal:
\red{\begin{eqnarray}
&&\forall x y(\nit{Ab}_R(x,y) \rightarrow \nit{End}_R(x,y)), \ \  \ \forall x y(\nit{End}_R(x,y) \rightarrow R(x,y)), \label{eq:end1}\\
&&~~\forall x(\nit{Ab}_S(x) \rightarrow \nit{End}_S(x)),  \ \ \ \ \ \ \ \ \ \ \ \ \forall x(\nit{End}_S(x) \rightarrow S(x)). \label{eq:end2}
\end{eqnarray}}

In addition to this specification, we have the observation $\mc{Q}$:
\begin{eqnarray}
\red{\exists x \exists y ( S(x) \land R(x, y) \land S(y)).} \label{eq:quer}
\end{eqnarray}
\end{enumerate}

Finally, we make the assumption that there are not abnormal tuples:
\red{\begin{eqnarray}
\forall x y(\nit{Ab}_R(x,y) \rightarrow \mbox{\bf false}), \ \ \ \
\forall x(\nit{Ab}_S(x) \rightarrow \mbox{\bf false}). \label{eq:last}
\end{eqnarray}}
\red{The FO theory formed by (\ref{eq:one}) - (\ref{eq:last}) (more precisely, (\ref{eq:one}), (\ref{eq:tres}), (\ref{eq:disj}), (\ref{eq:quer}) and   (\ref{eq:last})) is inconsistent.}
\boxtheorem
\end{example}

Now, in more general terms, the observation is $\mc{Q}$ (being true), obtained by evaluating query $\mc{Q}$ on (theory of) $D$. In this case, $D \not \models \kappa(\mc{Q})$. Since all the abnormality predicates are assumed to
be empty, $\kappa(\mc{Q})$ is equivalent to $\kappa(\mc{Q})^\nit{Ab}$, which also becomes false with respect to $D$. As a consequence,  $\nit{SD} \cup \{(\ref{eq:default})\} \cup \{\mc{Q} \}$ is an inconsistent FO theory.
A
 diagnosis is a set of endogenous tuples that, by becoming abnormal, restore consistency.
 \begin{definition}\label{def:diag}
 \begin{enumerate}[(a)]
 \item A {\em diagnosis} for $\mc{M}$ is a $\Delta \subseteq D^n$, such that
\begin{equation*}\nit{SD} \ \cup \ \{\nit{Ab}_P(\bar{c})~|~P(\bar{c}) \in \Delta\} \ \cup \ \{\neg \nit{Ab}_P(\bar{c})~|~P(\bar{c}) \in D \smallsetminus \Delta\} \ \cup \ \{\mc{Q}\} \vspace{-7mm}
\end{equation*}
\phantom{po}\\ \noindent is consistent.
\item
$\nit{Diag}^s(\mc{M},t)$ denotes the set of S-minimal diagnoses for $\mc{M}$ that contain tuple $t \in D^n$.

\item $\nit{Diag}^c(\mc{M},t)$ denotes the set of  C-minimal diagnoses
in $\nit{Diag}^s(\mc{M},t)$. \ignore{$\mc{M}$ that contain a tuple $t \in D^n$ and have the minimum cardinality (among those diagnoses that contain $t$)} \boxtheorem
\end{enumerate}
\end{definition}

\begin{example}\label{ex:mbdaex5p5}   (ex. \ref{ex:mbdaex5} cont.)  The theory can be made consistent by giving up
(\ref{eq:last}), and  making S-minimal sets of
tuples abnormal. \red{According to (\ref{eq:end1})-(\ref{eq:end2}), those tuples have to be endogenous.}

$\mc{M}$ has two S-minimal diagnosis:
$\Delta_1=\{ S(a_3)\}$ and $\Delta_4=\{ S(a_4)\}$. The first one corresponds to replacing
the second formula in (\ref{eq:last}) by $\forall x(\nit{Ab}_S(x) \wedge x \neq a_3 \rightarrow \mbox{\bf false})$, obtaining now a consistent
theory.

Here, $\nit{Diag}^s(\mc{M},S(a_3))= \nit{Diag}^c(\mc{M}, S(a_3))=\{ \{ S(a_3)\}\}$, and $\nit{Diag}^s(\mc{M},$ $ S(a_4))=\nit{Diag}^c(\mc{M}, S(a_4))=\{\{ $ $S(a_4)\}\}$.

\red{If $R(a_4,a_3)$ is also endogenous, then also $\{R(a_4,a_3)\}$ becomes a minimal diagnosis.}
\boxtheorem
\end{example}

By definition,
$\nit{Diag}^c(\mc{M},t) \subseteq \nit{Diag}^s(\mc{M},t)$. Diagnoses for $\mc{M}$
and actual causes for $\mc{Q}$ are related.

\begin{proposition}\label{pro:ac&diag} \em
Consider  an instance $D$, a BCQ $\mc{Q}$, and the diagnosis problem $\mc{M}$ associated to $\mc{Q}$. Tuple $t \in D^n$ is an actual cause for $\mc{Q}$
iff $\nit{Diag}^s(\mc{M},t) \not = \emptyset$. \boxtheorem
\end{proposition}

The responsibility of an actual cause $t$ is determined by the cardinality of the diagnoses in $\nit{Diag}^c(\mc{M},t)$.
\begin{proposition}\label{pro:r&diag} \em
For  an instance $D$, a BCQ $\mc{Q}$, the associated diagnosis problem $\mc{M}$, and a tuple $t \in D^n$, it holds:

\noindent
(a) $\rho_{_{\!D\!}}(t)=0$ iff $\nit{Diag}^c(\mc{M},t) = \emptyset$.

\noindent  (b) Otherwise, $\rho_{_{\!D\!}}(t)=\frac{1}{|\Delta|}$, where $\Delta \in \nit{Diag}^c(\mc{M},t)$. \boxtheorem
\end{proposition}
 For the proofs of Propositions \ref{pro:ac&diag} and \ref{pro:r&diag}, it is easy to verify that the conflict sets of $\mc{M}$ coincide with the sets in $\mf{S}(D^n)$ (cf. Definition \ref{def:hsStuff}). The results are obtained from the characterization of minimal diagnosis as minimal hitting-sets of sets of conflict sets (cf. Section \ref{sec:prel} and \cite{Reiter87}) and Proposition \ref{pro:UBCQCauses}.
\begin{example}\label{ex:mbdaex6}   (ex. \ref{ex:mbdaex5p5} cont.)   From Propositions
\ref{pro:ac&diag} and \ref{pro:r&diag}, $S(a_3)$ and $S(a_4)$ are actual cases, with responsibility $1$. \red{If $R(a_4,a_3)$ is also endogenous, it also becomes an actual
cause with responsibility 1.}\boxtheorem
\end{example}

In consistency-based diagnosis, minimal
diagnoses can be obtained as S-minimal hitting-sets of the collection of S-minimal {\em conflict sets}  (cf. Section \ref{sec:prel}) \cite{Reiter87}.  In our case, conflict sets are S-minimal
sets of endogenous tuples  that, if not abnormal (only endogenous ones can be abnormal), and together, and possibly in combination with exogenous tuples, make (\ref{eq:ext}) false.

It is easy to verify that the conflict sets of $\mc{M}$ coincide with the sets in $\mf{S}(D^n)$ (cf. Definition \ref{def:hsStuff} and
Remark \ref{rem:hs}).  As a consequence, conflict sets for $\mc{M}$ can be computed in PTIME, the hitting-sets for $\mc{M}$ contain actual causes for $\mc{Q}$, and the hitting-set problem
for the diagnosis problems is of the $d$-hitting-set kind.

\red{The reduction from causality to consistency-based diagnosis allows us  to apply constructions and techniques for the latter (cf. \cite{Feldman10,Mozetic94}), to the former.}

\begin{example}\label{ex:mbdGrep}  (ex. \ref{ex:mbdaex5} cont.)
The diagnosis problem  $\mathcal{M}=( \nit{SD},\{S(a_4),S(a_3)\},$ $ \mc{Q})$ gives rise to the hitting-set framework $\mf{H}^n(D) = \langle  \{S(a_4),S(a_3)\}, \{\{ (S(a_3),$ $ S(a_4) \}\}\rangle$, with  $\{ S(a_3), S(a_4) \}$ corresponding to the
conflict set  $c=\{S(a_4),$ $S(a_3)\}$.

$\mf{H}^n(D)$ has two minimum hitting-sets: $\{ S(a_3) \}$ and  $\{S(a_4) \}$, which are the S-minimal diagnosis for $\mathcal{M}$. Then, the two tuples are actual causes for $\mc{Q}$ (cf. Proposition\  \ref{pro:ac&diag}). From Proposition \ref{pro:r&diag}, $\rho_{_{\!D\!}}(S(a_3))= \rho_{_{\!D\!}}(S(a_4))= 1$.  \boxtheorem
\end{example}

The solutions to the diagnosis problem can be used for computing repairs.
\begin{proposition}\label{pro:diag} \em
Consider an instance $D$ with $D^x = \emptyset$, a set of DCs of the form $\kappa\!: \ \forall   \bar{x}\neg (P_1(\bar{x}_1)  \wedge  \cdots \wedge
P_m(\bar{x}_m)$, and their associated ``abnormality-aware" integrity constraints\footnote{Notice that these are not denial constraints.}
in (\ref{eq:ext}) (in this case we do not need $\nit{End}_P$ atoms).

Each S-minimal diagnosis $\Delta$ gives rise to an S-repair of $D$, namely $D_{\!\Delta} = D \smallsetminus \{P(\bar{c}) \in D~|~
\nit{Ab}_P(\bar{c}) \in \Delta\}$;  and every S-repair can be obtained in this way. Similarly, for C-repairs using C-minimal diagnoses. \boxtheorem
\end{proposition}
\begin{example}\label{ex:mbdaex6+} (ex. \ref{ex:mbdaex6} cont.)
The instance  $D=\{S(a_3),$ $S(a_4),$ $R(a_4,a_3) \}$, \red{with all tuples endogenous,} has three (both S- and C-) repairs  with respect to the DC $\kappa\!: \ \forall x y \neg ( S(x)  \land  R(x, y) \land  S(y))$, namely \red{$D_1 = \{S(a_3),R(a_4,a_3) \}$,
$D_2 = \{S(a_4),R(a_4,a_3) \}$, and $D_3 = \{S(a_3),S(a_4)\}$.} They can be obtained as $D_{\!\Delta_1}, D_{\!\Delta_2},$ $ D_{\!\Delta_3}$ from the only (S- and C-) diagnoses,  $\Delta_1=\{ S(a_3)\}$, $\Delta_2=\{ S(a_4)\}$, $\Delta_3=\{R(a_4,a_3)\}$, resp.
\boxtheorem
\end{example}

\red{We have characterized repairs in terms of diagnosis. Thinking of the other direction, and
as a final remark, it is worth observing that the very particular kind of diagnosis problem we introduced above (with restricted logical formulas) can be formulated as a {\em preferred-repair problem} \cite[Sec. 2.5]{2011Bertossi}. Without going into the details, the idea is to
materialize tables for the auxiliary predicates $\nit{Ab}_P$ and $\nit{End}_P$, and consider the DCs of the form  (\ref{eq:ext}) (with the $\nit{End}_P$ atoms when not all
tuples are endogenous), plus the DCs (\ref{eq:default}), saying that the initial extensions for the $\nit{Ab}_P$ predicates are empty.}
If $D$ is inconsistent with respect to this set of DCs, the S-repairs that are obtained by only {\em inserting} endogenous tuples into the
extensions of the $\nit{Ab}_P$ predicates correspond to S-minimal diagnosis, and each S-minimal diagnosis can be obtained in this way.

\section{ \ Complexity Results}\label{sec:MBDcomx}

There are {\em three main computational problems} in database causality. For a BCQ $\mc{Q}$ and database $D$:

\begin{itemize}
\item[(a)] \ The {\em causality problem} (CP) is about computing the actual causes for $\mc{Q}$. Its decision version of this problem,
CDP,
is stated in (\ref{eq:cpd}).
\ Both CP and CDP
are solvable in polynomial time \cite{Meliou2010a}, which can be extended to UBCQs (cf. Proposition \ref{pro:UBCQCausesindirect}).
\item[(b)] \ The
{\em responsibility problem} (RP) is about computing the responsibility $\rho_{_{\!D\!}}(t)$ of a given actual cause $t$. \ (Since a tuple that is not
an actual cause has responsibility $0$, this problem subsumes (a).) \ \red{This is a maximization
problem due to the minimization of $|\Gamma|$ in the denominator.}

\red{We will consider the  decision version of this problem that, as usual for maximization problems \cite{garey}, asks whether the real-valued function being computed
(responsibility in this case)
takes a value greater than a given threshold $v$ of the form $\frac{1}{k}$, for a positive  integer $k$.}
\end{itemize}

\begin{definition}  \label{def:resp}   For  a BCQ $\mc{Q}$,
the {\em responsibility decision problem} (RDP) is (deciding about membership of):

\vspace{1mm}
$\mathcal{RDP}(\mc{Q})=\{(D,t,v)~|~t \in D^n, v \in \{0\} \cup \{\frac{1}{k}~|~k \in \mathbb{N}^+\}, \mbox{ and}$

\vspace{1mm}\hspace*{7.6cm}$D \models \mc{Q} \ \mbox{ and } \ \rho_{_{\!D\!}}(t) > v  \}$,\\
that is, deciding if a tuple has a responsibility greater than a bound $v$ (as a cause for $\mc{Q}$). \boxtheorem
\end{definition}
\red{The complexity analysis of RDP in \cite{Meliou2010a} is restricted to conjunctive queries without self-joins.} Here, we will generalize the complexity analysis for RDP to general CQs.

\vspace{1mm}
\noindent (c) \ Computing the {\em most responsible actual causes} (MRC). \
Its decision version, MRCDP, the {\em most responsible cause decision problem}, is a natural problem, because actual causes with the highest responsibility tend
to provide most interesting explanations for query answers
\cite{Meliou2010a,Meliou2010b}.  
\begin{definition}  \label{def:mracp}   For a BCQ $\mc{Q}$, the {\em most responsible cause decision problem}
 is (membership of):

\vspace{1mm}
 $\mc{MRCDP}(\mc{Q})$ $=\{(D,t)~|~ t \in D^n \mbox{ and } 0 < \rho_{_{\!D\!}}(t) \mbox{ is a maximum for}
 \ D\}$.
\boxtheorem\\
\end{definition}

We start by analyzing a more basic decision problem, that of deciding if a set of tuples $\Gamma$ is an S-minimal contingency
set associated to a cause $t$ (cf.  (\ref{eq:ct})).
\ignore{\begin{definition}  \label{def:cusp}  For a BCQ $\mc{Q}$,  the {\em minimal contingency set decision problem} (MCSDP) is
\vspace{1mm}
$\mc{MCSDP}(\mc{Q}):=\{(D,t,\Gamma)~|~\Gamma \in \nit{Cont}(D,\mc{Q},t)\}$.
\boxtheorem
\end{definition} }
Due to the results in Sections \ref{sec:causfrepair} and \ref{sec:repairfcauses}, it is clear that there is a close connection between this problem and  the {\em S-repair checking} problem \cite[Chap. 5]{2011Bertossi},
 about deciding if instance $D'$ is an S-repair of instance $D$ with respect to a set of integrity constraints. Actually,
the following result is obtained from the PTIME solvability of the S-repair checking problem for DCs \cite{Chomicki05} (see also \cite{Afrati09}).
\red{\begin{proposition}\label{pro:CSPCcpx} \em For  a BCQ $\mc{Q}$, the {\em minimal contingency set decision problem} (MCSDP), \ i.e.
\ \
$\mc{MCSDP}(\mc{Q}):=\{(D,t,\Gamma)~|~\Gamma  \mbox{ is minimal element in}$ $\nit{Cont}(D,\mc{Q},t)\}$, \ belongs to $\nit{PTIME}$. 
\end{proposition}}
\proof{To decide if $(D,t,\Gamma) \in  \mc{MCSDP}(\mc{Q})$, it is good enough to observe, from  Proposition \ref{pro:c&r},  that $(D,t,\Gamma) \in  \mc{MCSDP}(\mc{Q})$ iff $D \smallsetminus  (  \Gamma \cup \{t\})$ is an S-repair for $D$ with respect to $\kappa(\mc{Q})$.  S-repair checking can be done in PTIME in data \cite{Chomicki05}.\boxtheorem\\}

We could also consider the decision problem defined  in Proposition \ref{pro:CSPCcpx}, but with C-minimal $\Gamma$. We will not
use results about this problem in the following. Furthermore, its connection with the C-repair checking problem is less direct. As one can see from Section \ref{sec:causfrepair},
 C-minimal contingency sets correspond to a repair semantics somewhere between the S-minimal and C-minimal repair semantics (a subclass of \nit{Srep}, but a superclass  of \nit{Crep}): \ It is about an S-minimal repair with minimum cardinality that does not contain a particular tuple.

Now we establish that RDP is \nit{NP}-complete for CQs in general. The \nit{NP}-hardness is shown in \cite{Meliou2010a}. Membership of \nit{NP}
is obtained using Proposition \ref{pro:CSPCcpx}.

\begin{theorem}\label{the:RP(D)cmx} \em
(a) For every BCQ $\mc{Q}$,  $\mathcal{RDP}(\mc{Q}) \in  \nit{NP}$.

\noindent  (b) \cite{Meliou2010a} \ There are CQs $\mc{Q}$ for which $\mathcal{RDP}(\mc{Q})$  is  \nit{NP}-hard. 
\end{theorem}

\proof{(a) We give
a non-deterministic PTIME algorithm to solve  RDP. Non-deterministically guess a subset $\Gamma \subseteq D^n$, return {\em yes} if $|\Gamma| < \frac{1}{v}$ and $(D, t,$ $ \Gamma ) \in  \mc{MCSDP}$; otherwise return {\em no}. According to Proposition \ref{pro:CSPCcpx} this can be done in PTIME in data complexity.\boxtheorem\\}

In order to better understand the complexity of RP, the responsibility computation problem, we will investigate the {\em functional}, non-decision version of RDP.

The main source of complexity when computing responsibilities is related to the hitting-set problem  associated to $\mf{H}^n(D) = \langle D^n, \mf{S}^n(D)\rangle$ in Remark \ref{rem:hs} (cf. (\ref{eq:frame})). In this case, it is about computing the cardinality of a minimum hitting-set that contains a given vertex (tuple) $t$. That  this is
a kind of {\em $d$-hitting-set problem} \cite{NiedermeierR03} will be useful in Section \ref{sec:fpt}.

\begin{remark}\label{rem:vcp}
Our responsibility problem  can also be seen
as a {\em vertex cover problem} on the {\em hypergraph}\footnote{\red{In an hypergraph $\mc{H}$, a  set of vertices is a {\em vertex cover} if it intersects every hyperedge. A minimal vertex cover has no proper subset that is also a vertex cover. A {\em minimum} vertex cover has minimum cardinality among the vertex covers. Similarly, an {\em independent set} of $\mc{H}$ is a set of vertices such that no pair of them is contained in a hyperedge. Maximal and maximum
independent sets are defined in an obvious manner.}}
\begin{equation}\label{eq:hyper}
\mf{G}^n(D) = \langle D^n, \mf{S}^n(D)\rangle
\end{equation}
 associated to $\mf{H}^n(D) = \langle D^n, \mf{S}^n(D)\rangle$ \ (that is, the hitting-set framework can be seen as a hypergraph). In it,
the  hyperedges are the members of $\mf{S}^n(D)$. Determining the responsibility of a tuple $t$ becomes the problem on hypergraphs of determining the size of a minimum
vertex cover that contains  vertex $t$ (among all vertex covers that contain the vertex).  Again, in this problem the hyperedges are
bounded in size by $|\mc{Q}|$.\footnote{We  recall that
repairs of databases with respect to DCs can be characterized as maximal independent sets of {\em conflict hypergraphs} ({\em conflict graphs} in the case of FDs) whose vertices are the database tuples, and hyperedges connect tuples
that together violate a DC \cite{Arenas03,Chomicki05}.} \boxtheorem
\end{remark}

\begin{example}\label{ex:hyperexm}  For $\mc{Q}\!: \exists xy(P(x)\wedge R(x,y)\wedge P(y))$, and $D= D^n=\{P(a), P(c),$ $ R(a,c),$  $R(a,a)\}$, \ $\mf{S}(D)=\mf{S}^n(D)=\{ \{P(a), R(a,a)\}, \{P(a), P(c), R(a,c)\} \}$.

\red{The hypergraph $\mf{G}^n(D)$ has $D$ as set of vertices, and its hyperedges are $\{P(a),$ $ R(a,a)\}$ and $\{P(a), P(c), R(a,c)\}$. Its minimal vertex covers are: $vc_1=\{ P(a)\}$, $vc_2=\{ P(c), $ $R(a,a)\}$, $vc_3=\{  R(a,a), R(a,c)\}$. Only the first  has minimum cardinality. Accordingly, its only element, $P(a)$,} is an actual cause with responsibility $1$. The other tuples are actual causes with responsibility $\frac{1}{2}$.
\boxtheorem
\end{example}

\begin{remark} \red{To simplify the presentation of the next computational problems (Lemmas \ref{lemma:resclx} and \ref{lemma:MMVCand} and Proposition \ref{pro:MMVCcml}), we will formulate and  address them in terms of graphs. However, they still hold for hypergraphs \cite{icdt07,icdt07ext}, which is what we need for the complexity results obtained in the rest of this section.}  \boxtheorem
\end{remark}


\begin{lemma}\label{lemma:resclx} \em {\em (representation lemma)} \
There is a fixed database schema $\mathcal{S}$ and a BCQ $\mc{Q} \in \mc{L}(\mathcal{S})$, without built-ins, such that, for every graph $G=(V,E)$, \red{with non-empty $E$}, and  $v \in V$, there is
an instance $D$ for $\mathcal{S}$ and a tuple $t \in D$, such that the size of a minimum vertex cover of $G$ containing
$v$ \red{is the inverse of the} responsibility of $t$ as an actual cause for $\mc{Q}$.
\end{lemma}

\proof{Consider a graph $G = (V, E)$, and assume the vertices
of $G$ are uniquely labeled.

Consider the database schema with relations
$\nit{Ver}(v_0)$ and $\nit{Edges}(v_1, v_2, e)$, and the conjunctive query $\mc{Q}\!: \exists v_1v_2e(\nit{Ver}(v_1) \wedge \nit{Ver} (v_2) \wedge \nit{Edges}(v_1, v_2, e))$.
$\nit{Ver}$ stores the vertices of $G$, and $\nit{Edges}$, the labeled edges. For each edge $(v_1, v_2) \in E$, $\nit{Edges}$ contains $n$ tuples of the form $(v_1, v_2, i)$, where $n$ is the
number of vertices in $G$. All the values in the third attribute of $\nit{Edges}$ are different, say from 1 to $n \times |E|$. \red{This padding of relation $\nit{Edge}$ will
 ensure in the rest of the proof that C-minimal contingency sets for the query answer consist only of vertices, i.e. elements of $\nit{Ver}$ (as opposed to \nit{Edge} tuples).} \
The size of the padded instance  is still polynomial in the size of $G$.  It is clear that $D \models \mc{Q}$.

Assume $\nit{VC}$ is the minimum vertex cover of $G$ that contains vertex $v$, \red{where tuple $t$ is $\nit{Ver}(v)$}.  Consider the set of tuples $\Lambda= \{ \nit{Ver}(x)~|~x \in \nit{VC} \}$. Since  $v \in \nit{VC}$,    $\Lambda=\Lambda' \cup \nit{\{Ver}(v)\}$. Then, $D \smallsetminus (\Lambda' \cup \nit{Ver}(v)) \not \models Q$. This is because for every tuple $\nit{Edge}(v_i, v_j,k)$ in the instance, either $v_i$ or $v_j$ belongs to $\nit{VC}$. Due to the minimality of $\nit{VC}$, $D  \smallsetminus \Lambda' \models \mc{Q}$.
Therefore,  tuple $\nit{Ver(v)}$ is an actual cause for $\mc{Q}$.

Suppose $\Gamma$ is  a C-minimal contingency set associated to  $\nit{Ver}(v)$. Due to the C-minimality  of
$\Gamma$, it entirely consists of tuples in $\nit{Ver}$. It holds that $D \smallsetminus (\Gamma \cup \{\nit{Ver}(v)\}) \not \models \mc{Q}$ and $D \smallsetminus \Gamma \models \mc{Q}$. Consider the set  $\nit{VC'}=\{x~|~ \nit{Ver}(x) \in \Gamma\} \cup \{v\}$. Since $D  \smallsetminus (\Gamma \cup \{\nit{Ver}(v)\}) \not \models \mc{Q}$, for every tuple $\nit{Edge}(v_i, v_j,k)$ in $D$, either $v_i \in \nit{VC}' $ or  $v_j \in  \nit{VC}'$. Therefore, $\nit{VC}'$ is a minimum vertex cover of $G$ that contains $v$. It holds that $\rho_{_{\!D\!}}(\nit{Ver}(v))=\frac{1}{1+|\Gamma|}$. So, the size of a minimum vertex cover of $G$ that contains $v$ can be obtained from $\rho_{_{\!D\!}}(\nit{Ver}(v))$.\boxtheorem\\}

Having represented our responsibility problem as a graph-theoretic problem, we first consider functional computational problems in graphs.

\begin{definition} The {\em minimal vertex cover membership problem} (MVCMP) consists in, given a graph $G=(V,E)$, and a  vertex $v \in V\!$ as inputs, computing the size of a minimum vertex cover of $G$ that contains $v$.
\boxtheorem
\end{definition}

\begin{lemma}\label{lemma:MMVCand} \em
Given a graph $G$ and a vertex $v$ in it,  there is a graph $G'$ extending $G$ that can be constructed in polynomial time in $|G|$, such that the size of
a minimum vertex cover for $G$ that contains $v$ and the size  of a minimum vertex cover for $G'$ coincide.
\end{lemma}

  \proof{The size of  $\nit{VC_G(v)}$, the minimum vertex cover of $G$ that contains the vertex $v$, can be computed from the size of $I_G$, the maximum independent set of $G$, that {\em does not} contain $v$. In fact,
  \begin{equation}\label{one}
  |\nit{VC_G(v)}|=|G|-|I_G|.
  \end{equation}
Since $I_G$ is a maximum independent set that does not contain $v$, it must contain
one of the adjacent vertices of $v$ (otherwise, $I_G$ is not maximum, and $v$ can be added to $I_G$). Therefore, $|\nit{VC_G(v)}|$ can be computed from the size of a maximum independent set $I$ that contains $v'$, one of the adjacent vertices of $v$.

Given a graph $G$ and a vertex $v'$ in it, a graph $G'$ that extends $G$ can be constructed in polynomial time in the size of $G$, \red{in such a way that: \ there is a maximum independent set $I$ of $G$ containing $v'$ iff $v'$ belongs to
every maximum independent set of $G'$ iff the sizes of maximum independent sets for $G$ and $G'$ differ by one.
\ Actually, graph $G'$  can be obtained by adding a new vertex $v''$ that is connected only to the neighbors of $v'$. It holds:\footnote{This construction
is inspired by \cite[Lemma 1]{icdt07}. More details can be found in \cite{icdt07ext}.}
\begin{eqnarray}
|I_G|&=&|I_{G'}|-1, \label{two}\\
|I_{G'}|&=&|G'|-|\nit{VC}_{G'}|, \label{three}
\end{eqnarray}
where $I_{G'}$ is a maximum indent set in $G'$, and $\nit{VC}_{G'}$ is a minimum vertex cover of $G'$.}
From (\ref{one}), (\ref{two}) and (\ref{three}), we obtain:  $|\nit{VC_G(v)}|= |\nit{VC}_{G'}|$.\boxtheorem\\}

From Lemma \ref{lemma:MMVCand} and the $\nit{FP}^{\nit{NP(log} (n))}$-completeness of determining the size of a maximum clique in a graph  \cite{Krentel88}, we obtain:

\begin{proposition}\label{pro:MMVCcml} \em
The MVCMP problem for graphs is $\nit{FP}^{\nit{NP(log} (n))}$-complete.
\end{proposition}

 \proof{We prove membership by describing
an algorithm in $\nit{FP}^{\nit{NP(log} (n))}$ for computing the size of the minimum vertex cover of a graph $G=(V,E)$ that contains a vertex $v \in V\!$.  We use  Lemma \ref{lemma:MMVCand}, and build the extended graph $G'$.

The size of a minimum vertex cover for $G'$ gives the size of the minimum vertex cover of $G$ that contains $v$.  Since computing the maximum cardinality of a clique can be done in time $\nit{FP}^{\nit{NP(log} (n))}$  \cite{Krentel88}, computing a minimum vertex cover  can be done in the same time (just consider the complement graph).  Therefore, MVCMP belong to $FP^{NP(log (n))}$.

Hardness can be obtained by a reduction from computing minimum vertex covers in graphs to MVCMP. Given a graph $G$ construct the graph $G'$ as follows:  Add a vertex $v$ to $G$ and connect it to all vertices of $G$. It is easy to see that $v$ belongs to all minimum vertex covers of
$G'$. Furthermore,  the sizes of minimum vertex covers for $G$ and $G'$ differ by one. Consequently, the size of a minimum vertex cover of $G$
 can be obtained from the size of a minimum vertex cover of $G'$ that contains $v$. Computing the minimum vertex cover is $\nit{FP}^{\nit{NP(log} (n))}$-complete. This follows from the $\nit{FP}^{\nit{NP(log} (n))}$-completeness of computing the maximum cardinality of a clique in a graph  \cite{Krentel88}.\boxtheorem\\}

\begin{theorem}\label{the:r&diag} \em
\begin{enumerate}[(a)] \item
 For every BCQ, $\mc{Q}$, computing the responsibility of a  tuple as a cause for $\mc{Q}$ is in $\nit{FP}^{\nit{NP(log} (n))}\!\!\!$.

\item
There is~a database schema and a BCQ $\mc{Q}$, without built-ins, such that computing the responsibility of a tuple as a cause for $\mc{Q}$ is $\nit{FP}^{\nit{NP(log} (n))}$-complete.
\end{enumerate}
\end{theorem}

\proof{\red{For membership, we observe from Remark \ref{rem:vcp} that computing a tuple's responsibility amounts to computing the size of a minimum vertex cover containing the tuple
in the graph associated to the query and instance at hand. By Proposition \ref{pro:MMVCcml}, this problem belongs to $\nit{FP}^{\nit{NP(log} (n))}$.}

\red{Hardness follows from Lemma \ref{lemma:resclx} and the hardness result in Proposition \ref{pro:MMVCcml}.} \boxtheorem}

Now we address the most responsible causes problem, MRCDP (cf. Definition \ref{def:mracp}). We use the connection with consistent query answering of Section \ref{sec:cqa}, namely Corollary \ref{cor:cqa&cox}, and the
$P^{\nit NP(log(n))}$-completeness of consistent query answering under the C-repair semantics for  queries that are conjunctions of ground atoms and a particular DC  \cite[Theorem 4]{icdt07}.

\begin{theorem}\label{the:cqa&ca&cox} \em
\begin{enumerate}[(a)] \item  For every BCQ, $\mc{MRCDP}(\mc{Q}) \in P^\nit{NP(log(n))}\!\!$. \item
There is a database schema and a BCQ $\mc{Q}$, without built-ins, for which $\mc{MRCDP}(\mc{Q})$ is $P^\nit{NP(log(n))}$-complete.
\end{enumerate}
\end{theorem}

\proof{(a)  To show that $\mc{MRCDP}(\mc{Q})$ belongs to $\nit{P}^{\nit{NP(log} (n))}$, consider first the hitting-set framework $\mf{H}^n(D) = \langle D^n, \mf{S}^n(D)\rangle$ (cf. Definition \ref{def:hsStuff} and  \ref{eq:frame}) and its associated hypergraph $\mf{G}^n(D)$ (cf. (\ref{eq:hyper})).

It holds that $t$ is a most responsible cause for $\mc{Q}$ iff   $\mf{H}^n(D)$ has a C-minimal hitting-set that contains $t$ (cf. Proposition \ref{pro:UBCQCauses}). Therefore, $t$ is a most responsible cause for $\mc{Q}$ iff $t$ belongs to some minimum vertex cover of $\mf{G}^n(D)$.

It is easy to see that $\mf{G}^n(D)$ has a minimum vertex cover that contains $t$  iff $\mf{G}^n(D)$ has a maximum independent set that does not contains $t$. Checking if $t$ belongs to all maximum independent set of $\mf{G}^n(D)$ can be done in  $P^\nit{NP(log(n))}$ \cite[Lemma 2]{icdt07}.

If $t$ belongs to all independent sets of $\mf{G}^n(D)$, then $(D,t) \not \in \mc{MRCDP}(\mc{Q})$; otherwise $(D,t) \in \mc{MRCDP}(\mc{Q})$. As a consequence, the decision can be made in time $P^\nit{NP(log(n))}$.

\vspace{1mm}
\noindent (b) The proof is by a reduction, via Corollary \ref{cor:cqa&cox}, from consistent query answering under the C-repair semantics for queries that are conjunctions of ground atoms, which was proved to be $P^\nit{NP(log(n))}$-complete in  \cite[Theorem 4]{icdt07}. Actually, that proof (of hardness) uses a particular database schema $\mc{S}$ and a DC $\kappa$.  In our case, we can use the same schema $\mc{S}$ and the violation query $V^\kappa$ associated to $\kappa$ (cf. Section  \ref{sec:repairfcauses}).\boxtheorem\\}

From  Proposition \ref{pro:UBCQCauses} and the $\nit{FP}^\nit{NP(log(n))}$-completeness of determining the size of  C-repairs for DCs \cite[Theorem 3]{icdt07}, we obtain the following for the computation of the highest responsibility value.

\ignore{
\combabak{We need the result from  \cite[theo. 3]{icdt07}, its not obtained directly from what we have so far. We need to change Lemma 2 to obtain it directly. }
\comlb{We should briefly explain why not.}
\combabak{ I am not sure what would you expect me to say here. How can I explain that I did not want to redo thing that already has been done.}
\comlb{NEW: I am not asking to explain that, but why it does not follow from that Lemma.}
\combabak{NEW: We may need to talk about this. I really don't know how to explain}       }

\begin{proposition}\label{pro:crepair&res&cox} \em \begin{enumerate}[(a)]\item  For every BCQ, computing the responsibility of the most responsible causes   is in $\nit{FP}^\nit{NP(log(n))}$.

\item
There is a database schema and a BCQ $\mc{Q}$, without built-ins, for which computing the responsibility of the most responsible causes is  $\nit{FP}^\nit{NP(log(n))}$-complete. 
\end{enumerate}
\end{proposition}

\proof{(a) To show the membership of $\nit{FP}^{\nit{NP(log} (n))}$, consider the hypergraph $\mf{G}^n(D)$ as obtained in Theorem \ref{the:cqa&ca&cox}. The responsibility of most responsible causes for $\mc{Q}$ can be obtained from the size of the minimum vertex cover of $\mf{G}^n(D)$ (cf. Proposition \ref{pro:UBCQCauses}).  The size of the minimum vertex cover in a graph can be computed in $\nit{FP}^{\nit{NP(log} (n))}$, which is obtained from the membership of $\nit{FP}^{\nit{NP(log} (n))}$ of computing  the maximum cardinality of a clique  in graph \cite{Krentel88}.

It is easy to verify that minimum vertex covers in hypergraphs can be computed in the same time.

\vspace{1mm}
\noindent (b) This is by a reduction from the problem of determining the size of C-repairs for DCs  shown to be  $\nit{FP}^\nit{NP(log(n))}$-complete in \cite[Theorem 3]{icdt07}. Actually, that proof (of hardness) uses a particular database schema $\mc{S}$ and a DC $\kappa$. In our case, we may consider the same schema $\mc{S}$ and the violation query
$V^{\kappa}$ associated to $\kappa$ (cf. Section  \ref{sec:repairfcauses}).

The size of C-repairs for an inconsistent instance $D$ of the schema $\mc{S}$ with respect to $\kappa$ can be obtained from the responsibility of  most responsible causes for $V^{\kappa}$ (cf.  Corollary  \ref{col:sr&cp}). \boxtheorem\\}

\subsection{ \ FPT of responsibility}\label{sec:fpt}

 We need to cope with the intractability of computing most responsible causes.
The area of {\em fixed parameter tractability} (FPT) \cite{flum} provides tools to attack this problem. In this regard, we recall that a decision problem with inputs
of the form $(I, p)$, where $p$ is a distinguished
parameter of the input, is fixed parameter tractable (or belongs to
the class FPT), if it can be solved in time $O(f(|p|) \cdot |I|^c)$, where $c$ and the
hidden constant do not depend on $|p|$ or $|I|$, and $f$ does not depend on $|I|$.

In our case, the {\em parameterized version of the decision problem} $\mathcal{RDP}(\mc{Q})$ (cf. Definition \ref{def:resp}) is denoted with $\mathcal{RDP}^p(\mc{Q})$, and the distinguished parameter is $k$, such
that
$v = \frac{1}{k}$.

That $\mathcal{RDP}^p(\mc{Q})$ belongs to FPT can be obtained from its formulation as a $d$-hitting-set problem ($d$ being the fixed upper bound on the size of the
sets in the set class). \red{The latter problem consists in, given a hitting-set framework with $d$-bounded subsets and an element $t$ (a tuple in our case),
deciding if there is a hitting-set of cardinality smaller that $k$ that contains $t$.} This problem  belongs to FPT.
\begin{theorem}\label{the:FPT} \em
For every BCQ $\mc{Q}$, $\mathcal{RDP}^p(\mc{Q})$ belongs to FPT, where the parameter is the inverse of the responsibility bound.
\end{theorem}

\proof{First, there is a  PTIME
parameterized algorithm for the $d$-hitting-set problem about deciding if there is a hitting-set of size at most  $k$ that runs in time
$O(e^k + n)$, with $n$ the size of the underlying set and  $e=d-1+o(d^{-1})$  \cite{NiedermeierR03}. In our case,  $n=|D|$, and $d = |\mc{Q}|$ \ (cf. also \cite{Fernau10}).

Now, to decide if the responsibility of a given tuple $t$ is greater than $v=\frac{1}{k}$, we consider the associated hypergraph $\mf{G}^n(D)$, and we decide
if it has a vertex cover that contains $t$ and whose size is less than $k$. In order to answer this, we use Lemma \ref{lemma:MMVCand}, and  build the extended hypergraph $\mf{G}'$.

The size of a minimum vertex cover for $\mf{G}'$
gives the size of the
minimum vertex cover of $\mf{G}^n(D)$ that contains $t$. If $\mf{G}^n(D)$ has a  vertex cover that contains $t$ of size less than $k$, then $\mf{G}'$ has a vertex cover of size less than $k$. If  $\mf{G}'$ has a vertex cover of size less than $k$,  its minimum size for a vertex cover is less than $k$. Since this minimum is the same as the size of a minimum vertex cover for $\mf{G}^n(D)$ that contains $t$, $\mf{G}^n(D)$ has a vertex cover of size less than $k$ that contains $t$. As a consequence,
it is good enough to decide if $\mf{G}'$ has a vertex cover of size less than $k$. For this, we use the hitting-set formulation of this hypergraph problem, and the already mentioned FPT algorithm.\boxtheorem\\}

This result and the corresponding algorithm sketched in its proof show  that the higher the required responsibility degree, the lower the  computational effort needed to compute the actual causes with at least that level of responsibility. In other
terms, parameterized algorithms are effective for computing actual causes with high responsibility or most responsible causes. In general, parameterized algorithms are
very effective when the parameter is relatively small \cite{flum}.

Now, in order to compute most responsible causes, we could apply, for each actual cause $t$, the just presented FPT algorithm on the hypergraph $\mf{G}^n(D)$, starting with $k=1$, i.e. asking if there is
  vertex cover of size less than $1$ that contains $t$. If the algorithm returns a positive result, then $t$  is a counterfactual cause, and has responsibility $1$. Otherwise, the algorithm will be launched with $k=2 , 3, \ldots, |D^n|$, until a positive result is returned. (The procedure can be improved through binary search on  $k=1, 2 , 3, \ldots,  m$, with $m$ possibly much smaller than $|D|$.)

The complexity results and algorithms provided in this section can be extend to UBCQs. This is due to Remark \ref{rem:ucq} and the construction of  $\mf{S}^n(D)$, which the results in this section build upon.  

For the $d$-hitting-set problem there are also efficient parameterized approximation algorithms \cite{Brankovic12}. They could be used to approximate the responsibility problem.
Furthermore,  approximation algorithms developed for the minimum vertex cover
problem on bounded hypergraphs \cite{Halperin00,Okun05} should be applicable to approximate most responsible causes for query answers. Via the causality/repair connection (cf. Section \ref{sec:cqa}), it should be possible to
develop approximation algorithms to compute  S-repairs of particular sizes, C-repairs, and consistent query answers with respect to DCs.

\subsection{ \ \red{Complexity of  diagnosis with positive disjunctive rules}}\label{sec:complMBD}

It is known that consistency-based diagnosis decision problems can be unsolvable \cite{Reiter87}.
However, there are decidable classes of FO diagnosis specifications, and those classes are amenable
to complexity analysis. However, there is little research on the complexity analysis of solvable classes
of consistency-based diagnosis problems.  The connection we established in the previous sections between
causality, repairs and
consistency-based diagnosis can be used to obtain new algorithmic and complexity results for
the latter. Without trying to be exhaustive about this, which is beyond the scope of this paper,
we give an example of the kind of results that can be obtained.

Considering the  diagnosis problem we obtained in Section \ref{sec:MBDtoRep}, we can define
a class of diagnosis problems. Cf. Example \ref{ex:mbdaex5}, in particular (\ref{eq:disj}), for
motivation.

\begin{definition}
A {\em disjunctive positive} (DP) diagnosis specification $\Sigma$ is a consistent FO logical theory, such that:
 \begin{enumerate}[(a)]
\item $\Sigma$ has a  signature (schema) consisting of a finite set of constants, a set of predicates $\mc{S}$, a set $\mc{S}^\nit{ab}$ of predicates of the form $\nit{Ab}_{\!R}$,\footnote{Or any other ``abducible"
predicates that are different from those in $\mc{S}$.}
with $R \in \mc{S}$, and $\nit{Ab}_{\!R}$ with the same arity of $R$. \ $\mc{S}$ and $\mc{S}^\nit{ab}$ are mutually disjoint.
\item $\Sigma$ is inconsistent with $\nit{AB}^\mc{S} := \{\forall \bar{x}(\nit{Ab}_{\!R}(\bar{x}) \rightarrow \mathbf{false})~|~R \in \mc{S}\}$.
\item Consists
of:
\begin{itemize}
\item [(c1)] Sentences of the form \ $\forall \bar{x}(C(\bar{x}) \ \longrightarrow \ \bigvee_i \! \nit{Ab}_{\!R_i}(\bar{x}_i))$, with
$\bar{x}_i \subseteq \bar{x}$, and $C(\bar{x})$ a conjunction of atoms that does not include $\nit{Ab}$-atoms of any kind.
\item [(c2)] Sentences of the forms \ $\forall \bar{x}(\!\nit{Ab}_{\!R}(\bar{x})\ \longrightarrow \ (R(\bar{x}) \wedge S(\bar{x})))$, with $S \in \mc{S}$.
\item [(c3)] A finite background universal theory $\mc{T}$ expressed in terms of predicates in $\mc{S}$ (and constants) that  has a unique Herbrand model.\footnote{This condition is clearly satisfied
by the logical reconstruction of a relational database, but can be relaxed in several ways.}   \boxtheorem
\end{itemize}
\end{enumerate}
\end{definition}
\ignore{Notice that in theory like this, the sentences
In this section we identify two computational problems in the context of
consistency-based diagnosis. Then we introduce a syntactic fragment of
first order theories for which we can provide some complexity results for
the corresponding problems in consistency-based diagnoses taking advantage
of the results we obtained in this paper.}

As above, a diagnosis is a set of $\nit{Ab}_{\!R}$-atoms that, when assumed to be true,
restores the consistency of the correspondingly modified $\Sigma \cup \nit{AB}^\mc{S}$.

There are at least two important computational tasks that emerge, namely, given a
{\em disjunctive positive} (DP) diagnosis specification $\Sigma$ together with $\nit{AB}^\mc{S}$:
\begin{enumerate}
\item The {\em minimum-cardinality diagnosis} (MCD) problem, about
computing \linebreak minimum-cardinality diagnoses.

 \item The {\em  minimal membership diagnosis}, (MMD) about
computing minimum-cardinality diagnoses that contain a given
\nit{Ab}-atom.
\end{enumerate}

It is not difficult to see that these problems are computable (or solvable in their decision
versions). Now we can obtain complexity lower bounds for them. Actually, in Section \ref{sec:MBDtoRep},
the {\em responsibility} and {\em most responsible
causes problem} were reduced to diagnosis problems for specifications that
turned out to be disjunctive positive (see (\ref{eq:disj})).

More specifically, Proposition \ref{pro:r&diag} reduces computing responsibility of a tuple to computing the size of a minimum-cardinality diagnosis that contains the tuple. Furthermore,  as a simple corollary of Proposition \ref{pro:r&diag}, we obtain the computation
of minimum-cardinality diagnoses allows us to compute most responsible causes.
Now, combining all this with Proposition
\ref{pro:crepair&res&cox} and  Theorem \ref{the:r&diag}, we obtain  the following lower bounds for our diagnosis problems.
\begin{theorem}\label{thm:diag} \em
For disjunctive  positive diagnosis specifications, the MCD and MMD problems
are $\nit{FP}^{\nit{NP(log} (n))}$-hard in the size of their underlying Herbrand structure. \boxtheorem
\end{theorem}

\section{ \ Preferred Causes for Query Answers}\label{sec:pref-cause}

   In Section \ref{sec:causfrepair} we characterized causes and most responsible causes in terms of S-repairs and C-repairs, resp. We could generalize the notion of a cause and/or its responsibility
   by using, in principle, any {\em repair semantics} $\mbox{\large \sf S}$. The latter is represented by a class of repairs \red{$\nit{Rep}^{\!\mbox{\large \sf S}\!}(D,\Sigma)$}, of $D$ with respect to a set of denial constraints (cf. Section \ref{sec:reps}).
 \ignore{  This class contains consistent instances over the same schema
   as $D$, and satisfy additional conditions. Actually, a repair semantics is based on two elements that determine $\nit{Rep}^{\!\mbox{\large \sf S}\!}(D,\Sigma)$: (a) the class of
admissible ``repair actions" (updates to restore consistency), and (b) a form of minimality condition that forces (minimal) repairs to stay as close as possible to the original instance $D$ \cite[sec. 2.5]{2011Bertossi}.}
When dealing with (sets of) DCs, the repair actions can only be of certain kinds. Usually  tuple deletions have been considered. This  is the case of the S- and C-repairs we have considered in this work
so far.

We could go beyond and consider  the notion of {\em prioritized repair} \cite{chomicki12}.
Also changes of attribute values can be the chosen repair actions, including the use of {\em null values}, to ``destroy" joins (again, with different semantics,
e.g. with nulls {\em \`a la} SQL \cite{iidb06,lechen}).

In this section we explore the possibility of introducing a notion of {\em preferred cause} that is based on a given repair semantics.
This idea is inspired by (and generalizes) the characterization of causes in terms of repairs that we obtained before, namely (\ref{eq:df}), (\ref{eq:dfc}), Proposition
   \ref{pro:c&r}, and Corollary \ref{cor:card}.

If we define causes and their (minimal) contingency sets on the basis of a given repair semantics, the minimality condition involved in the latter will have an impact on
the notion of minimal (or preferred) contingency set, and indirectly, on the notions of responsibility and most responsible cause.\footnote{We could say that the efforts in
\cite{halpern14,halpern15} to modify the Halpern-Pearl (HP) original definition of causality are  about considering more appropriate restrictions on contingencies. Since in some cases
the original HP definition does not provide intuitive results regarding causality, the modifications avoid this by recognizing some contingencies as ``unreasonable" or ``farfetched".}

\ignore{
\comlb{Are you sure the discussion by  Halpern in his last papers is about playing with different notions of (minimal) contingency?}
\combabak{ The main issue with the HP definition is ``what kind of restrictions musts be considered over contingencies  in general?" Actually, the original definition produced non-sense causes in some scenarios (concrete examples), because
it allows some contingencies that are not plausible (with respect to the mentioned  scenarios). The question is what contingencies (possible worlds) are plausible or conceivable in general. Halpern try to provide this generic condition that can deal with all scenarios and counter examples.}
\combabak{
The relevance of the previous comment to our problem: we can borrow the fact that given a scenario some contingencies are less proffered (less plausible) than others. For instance, when a user prefer a table $P$ to be cause (meaning that actual causes coming from $P$) of an unexpected query answer implicitly she assumed that possible states of the database (obtained by removing contingency sets) that does not contain the tuples from $P$ are more conciliable or plausible. Same discussion holds for preferred repairs (or any possible-world-based notions in databases e.g. view-updates)}
}

In Section \ref{sec:pref-rep} we summarize prioritized repairs. In Section \ref{sec:prefc-rep} we impose preferences on causes on the basis of the prioritized repairs introduced in \cite{chomicki12} (and further investigated in
\cite{fagin15}). In Section \ref{sec:endo}, we briefly investigate the possibility of capturing endogenous repairs, i.e. that do not
change exogenous tuples, by means of a priority relation. Finally, in Section \ref{sec:null}, we briefly consider the possibility of defining (preferred) causes via attribute-based repairs that use null values.

\subsection{ \ Prioritized repairs}\label{sec:pref-rep}

    The prioritized repairs in \cite{chomicki12} are based on a {\em priority relation}, $\succ$, on the set of database tuples. In the case of a pair of (mutually) {\em conflicting tuples}, i.e. that simultaneously violate a constraint in a given set
    set of DCs (possibly in company of other tuples), the repair process  reflects the user preference -as captured by the priority relation- on the tuples that are privileged to be kept in the database, i.e. in the intended repairs.

    Given such a priority relation, in  \cite{chomicki12} different classes of prioritized repairs are introduced,  namely the class of  {\em globally optimal repairs}, that of {\em Pareto-optimal repairs}, and that
    of {\em completion-optimal repairs}. Intuitively, each class relies on a  different {\em optimality criterion} that is used  to extend the priority relation $\succ$ on pairs of conflicting facts to a priority relation on the set of S-repairs. As a consequence, each of these three classes is contained in that of the S-repairs. In particular, all these repairs are based on tuple deletions.

 \ignore{   \comlb{When you say S-repairs above, you include the maximal consistency condition?}
\combabak{thats right}   }

    Let us denote with $\nit{Rep}^{\succ,\mc{X}}(D,\Sigma)$ the  class  of all prioritized repairs based on $\succ$ and the optimality criterion $\mc{X}$. Its elements are called {\em $(\sem)$-prioritized repairs} of $D$ with respect to a the set $\Sigma$ of DCs. It holds $\nit{Rep}^{\succ,\mc{X}}(D,\Sigma) \subseteq \nit{Srep}(D,\Sigma)$, and then,
    all the elements of $\nit{Rep}^{\succ,\mc{X}}(D,\Sigma)$ are subsets of $D$.

  \ignore{\comlb{We should have an example of preferred repair. Hopefully the example can be used below, when connecting with priorities/weights. (You are assuming everybody knows what a
    preferred repair.) We should also say what kind of properties $\succ$ is always assumed to have. We do ot have to illustrate the three classes, only the most relevant of the one for which there
    are more interesting results available in the literature}
\red{ +++++++++++++++++++++++++ new stuff}  }

\ignore{
Generalizing (\ref{eq:df}), (\ref{eq:dfc}), Proposition
   \ref{pro:c&r}, and Corollary \ref{cor:card}, we define a general notion of preferred causes. Notice that the following only provides a syntactic framework to define a family of preferred causes. That is
   we do not make any assumptions on what kind of preferences we consider in causality. We do not even assume any connection between a preference relation  $\succ$ in the context of preferred repairs and causality.
\combabak{I used  $\succ'$ to show preference in causality as yet we have no understanding of the relation between a preference in repair and causality.}
\comlb{I do not see why: the preference on causes is based on that on repairs (or tuples).}
\boxtheorem
}

In order to show a concrete class $\nit{Rep}^{\succ,\mc{X}}(D,\Sigma)$, we first recall the  definitions of priority relation and {\em global-optimal repair} from \cite{chomicki12}.

\begin{definition} \label{def:PI}
Given an instance $D$ and a set of denial constraints $\Sigma$ , a binary relation $\succ$ on $D$ is a {\em priority relation} with respect to $\Sigma$ if: (a) $\succ$ is acyclic, and (b) for
every $t,t' \in D$, if $t  \succ t'$, then $t$ and $t'$  are mutually conflicting.\footnote{We can say $\{t,t'\}$ is a {\em conflict}, i.e.
the two tuples jointly participate in the violation of one of the DCs in $\Sigma$.}  \boxtheorem
\end{definition}

\begin{definition} \label{def:PR}
Let $D$ be an instance,  $\Sigma$ a  set of DCs, and $\succ$ a corresponding priority relation. Let  $D'$ and $D''$ be two consistent sub-instances of $D$.  $D'$ is  a {\em global improvement} of $D''$ if
$D' \not =D''$, and for every tuple  $t' \in D''\smallsetminus D'$, there exists a tuple $t \in D'\smallsetminus D''$ such that $t \succ t'$. $D'$ is a {\em global-optimal repair} of $D$, if $D'$ is an S-repair and
does not have a global improvement. \boxtheorem
\end{definition}
In  this definition, the optimality criterion, a possible $\mc{X}$ above, is that of global-optimal repair, or $(\succ\!,\nit{go})$-repair, which leads to a class $\nit{Rep}^{\succ,\nit{go}}(D,\Sigma)$.
We consider this repair semantics just for illustration purposes.
\begin{example}\label{ex:pfcex1} \
Consider the database  schema  $\nit{Author}(\nit{Name,Journal})$,\\ \hspace*{2cm} $\nit{Journal}(\nit{JournalN},$ $\nit{Topic,Paper\#})$, and the following instance $D$:

\vspace{0.3cm}

\hspace*{-3mm} \begin{tabular}{l|c|c|} \hline
Author & \nit{Name} & \nit{Journal} \\\hline
\ignore{ & Joe  & TKDE\\}

& John & TKDE\\
& Tom & TKDE\\
& John & TODS\\
\hhline{~--}
 \end{tabular} ~~~~\begin{tabular}{l|c|c|c|} \hline
Journal  & \nit{JournalN} &\nit{Paper\#} &  \nit{Topic}\\\hline
 & TKDE & 30 & XML\\
& TKDE &31 & CUBE\\
& TODS & 32 & XML\\
\hhline{~---}
\end{tabular}
 \\

 Consider the following denial constraint:
 \begin{equation}
 \kappa\!: \ \forall x y z z'  \neg (\nit{Author}(\nit{x,y}) \land \nit{Journal}(\nit{y,z,z'}) \land x={\sf John} \land z'={\sf XML}), \label{eq:kappa}
\end{equation}
capturing the condition that \red{``John has not published a paper  in a journal that has published papers on XML"}.

$D$ is inconsistent with respect to $\kappa$, and contains the following sets of conflicting tuples:
\begin{eqnarray}
C_1&= & \{\nit{Author(John,TKDE), Journal(TKDE,30,X\!\!ML)}\}, \nonumber \\
C_2&=&\{\nit{Author(John,TODS), Journal(TODS,32,X\!\!ML)}\}. \nonumber
\end{eqnarray}
$D$ has the following S-repairs, each  obtained by deleting one tuple from each of $C_1$ and $C_2$, to resolve the conflicts:
{\small
\begin{eqnarray*}
D_1&= & \{\nit{\red{Author(Tom,TKDE)},Journal(TKDE,31,CUBE),Author(John, TODS)},\\&&~~\nit{Journal(TKDE,30,XML)}\} \nonumber \\
D_2&= & \{\nit{Author(Tom,TKDE),Journal(TKDE,31,CUBE),Journal(TKDE,30,XML)},\\&&~~\nit{Journal(TODS,32,XML)}\} \nonumber \\
D_3&= & \{\nit{Author(Tom,TKDE),Journal(TKDE,31,CUBE),Author(John,TKDE)},\\&&~~\nit{Journal(TODS,32,XML)}\} \nonumber \\
D_4&= & \{\nit{Author(Tom,TKDE),Journal(TKDE,31,CUBE),Author(John,TKDE)},\\&&~~\nit{Author(John, TODS)}\} \nonumber
\end{eqnarray*}}
(a) \ Now, assume  a user prefers to resolve a conflict by removing tuples from the \nit{Author} table rather than the \nit{Journal} table, maybe because he considers the latter more reliable than the former.
This is expressed the following priority relationships on conflicting tuples: {\small \nit{Journal(TKDE,30,XML)} $\succ$ \nit{Author(John,TKDE})} and  {\small \nit{Journal(TODS,32,XML)} $\succ$ \nit{Author(John,TODS)}}.

In this case only $D_2$ is a global-optimal repair. Actually, $D_2$ is a  global improvement over each of $D_1$, $D_3$ and $D_4$. For $D_1$, for example: \red{$D_2 \smallsetminus D_1=\{${\small \nit{Journal(TODS,32,XML}}$\}$ and $D_1 \smallsetminus D_2=\{${\small \nit{Author(John, TODS)}}$\}$.} We can see that, for each tuple in $D_2 \smallsetminus D_1$, there is a tuple in $D_1 \smallsetminus D_2$ that has a  higher priority. Therefore, $D_2$ is a global improvement on $D_1$. So, in this case $\nit{Rep}^{\succ,\nit{go}}(D,\kappa) = \{D_2\}$

In this case, the uniqueness of the global-optimal repair is quite natural as the preference relation among conflicting tuples is a total relation. So, we know how to resolve every conflict according to the user preferences.

(b) \ For a more subtle situation, assume the user has the priorities as before, but in addition he tends to believe that John has a paper in TODS. In this case we have only the relationship {\small \nit{Journal(TKDE,30,XML)} $\succ'$ \nit{Author(John,TKDE)}},
and no preference for resolving the second conflict. Now both $D_1$ and $D_2$ are  global-optimal repairs. That is, now $\nit{Rep}^{\succ'\!\!,\nit{go}}(D,\kappa) = \{D_1, D_2\}$.
 \boxtheorem
\end{example}

\ignore{
In the context of causality the exogenous/endogenous partitioning  tuples reflects a user causality preference on a database tuples in a binary way
(a tuple is either exogenous or endogenous) provided by restricting causes to be among endogenous tuples.

Recall from Section \ref{causeandres} that checking weather a tuples is a cause is essentially done trough a (hypothetical) sequence of tuples deletions to
check of there exists a (hypothetical) state of database in which the tuple is a critical (counteract cause) for the answer. The underling
concept behind the exogenous/endogenus partitioning is that hypothetical states of a database that are obtained bye removing some exogenous tuples (possibly together with some endogenous tuples) are  ``unreasonable" or ``farfetched".

The idea can be generalize and applied to the context of database repairs (and possibly other problems in databases). To generalize the idea lets call a tuple $t$ an {\em assumable fact} if a ``conceivable" or ``plausible" hypothetical state of a database can be obtained through deletion of $t$ (and possibly other tuples). Oppositely, call $t$ a {\em non-assumable fact} if all the hypothetical states of the database obtained by removing $t$  (and possibly other tuples) are  ``unreasonable" or ``farfetched".

Now, given an inconsistent database $D$ partitioned into assumable and non-assumable with respect to a set of DC $\Sigma$. The {\em assumable minimal-repairs} of $D$ with respect to some minimality criteria are those minimal repairs that are subset of assumable tuples.
}

\subsection{ \ Preferred causes from prioritized repairs}\label{sec:prefc-rep}

According to the motivation provided at the beginning of this section, we now define {\em preferred causes} on the basis of a class of prioritized repairs. (Compare (\ref{eq:dif}) below with (\ref{eq:df}) and (\ref{eq:dfc}).)
To keep things simple, we concentrate on single BCQs, $\mc{Q}$, whose associated denial constraints are denoted by $\kappa(\mc{Q})$.

Before providing technical details, we motivate  the notion of preference in the context of causality. In this direction, first notice that under actual causality, we already  make a difference -and only this difference- between endogenous and exogenous tuples. We can think of extending this priority
relation among tuples in such a way that, for example, we prioritize -as causes- tuples in a given relation $R$, and we are not interested in tuples in another relation $S$.
\ignore{In the rest of this section we argue that attribution of preferred to actual causes in our proposal follows our intuition of a plausible account for  preferred causation. We naturally desire that preference in causality provides an extension to exogenous/endogenus partitioning which reflects a user preference in a binary way (a tuple is either exogenous or endogenous). We desire a definition of  preference in which instead of specifying for example that relation $R$ is exogenous and
$S$ endogenous, meaning that the user is only interested in causes in $S$,} So, the user can specify a priority relation between the two relations, or different {\em scores}  for these relations \cite{Meliou2011}.


 In Section \ref{sec:disjcausesCont} actual causes and their minimal contingency sets for a UBCQ were characterized as the minimal hitting-sets of the collection $\mc{C}$ of minimal subsets of a database that entail the query.
 Those minimal hitting-sets are obtained by removing at least one tuple from each of the elements of $\mc{C}$ (cf. Proposition \ref{pro:UBCQCauses}). At this point, user preferences, or priorities, could be applied to tuples that belong to a same set $\mc{C}$.

\begin{definition} \label{def:jc}
Given an instance $D$ and a BCQ $\mc{Q}$,  tuples $t$ and $t'$ are {\em jointly-contributing} if $t \not =t'$, and there exists an  S-minimal $\Lambda \subseteq D$ such that $\Lambda \models \mc{Q}$ and $t,t'  \in \Lambda$.
\boxtheorem
\end{definition}

Now we define priority relations on jointly-contributing tuples.
\begin{definition} \label{def:pricause}
Given an instance $D$ and a BCQ $\mc{Q}$, a binary relation $\succ_{\!c}$ on $D$ is a {\em causal priority relation} with respect to $\mc{Q}$ if: (a) $\succ_{\!c}$  is acyclic, and (b) for every $t,t' \in D$, if $t  \succ_{\!c} t'$, then $t$ and $t'$ are jointly-contributing tuples.\boxtheorem
\end{definition}

This definition introduces a natural notion of preference on causality. Actually, this way of approaching priorities on causes is in (inverse) correspondence with
 preference on repairs as based on priority relations on conflicting tuples. To see this, first observe that for a given instance $D$ and BCQ $\mc{Q}$: $t$ and $t'$ are jointly-contributing tuples  for $\mc{Q}$ \ iff \ $t$ and $t'$ are mutually conflicting tuples for $\kappa(\mc{Q})$.

 Next, in the context of prioritized repairs, a priority relation reflects a user preference on tuples that are preferred to be kept in the database. This is the inverse of  causality, where
 a causal priority relation, as we defined it, reflects the tuples that are preferred to be (hypothetically or counterfactually) removed from database, to make them preferred causes.

In the following assume $\succ_{\!c}^r$ is the inverse of a causal priority  relation $\succ_{\!c}$. That is, $t \succ_{\!c}^r t'$ \ iff \ $t' \succ_{\!c} t$. Clearly, $\succ_{\!c}^r$
 is acyclic, and can be imposed, with the expected result, on pairs of conflicting tuples. As a consequence, $\succ_{\!c}^r$ can be used to define prioritized repairs.

\begin{definition} \label{def:prefCuases}
   Let  $D$ be an instance, $\mc{Q}$  a BCQ, $t$ a tuple in $D$, $\succ_{\!c}$ a causal priority relation on $D$'s tuples.
\begin{eqnarray} \label{eq:dif}
 \hspace*{-.8cm}\mbox{(a)}  \ \nit{Diff}^\semcr(D,\kappa(\mc{Q}), t):=\{ D \smallsetminus D'&|& D' \in \nit{Rep}^{\semcr}(D,\kappa(\mc{Q})), \mbox{ and}\nonumber\\&& \hspace*{3cm} t \in D\smallsetminus D'\}.
\end{eqnarray}
\hspace*{.1cm}\mbox{(b)} \ $t \in D$ is a \ $(\semc)${\em -preferred cause} for $\mc{Q}$ \  iff \
$\nit{Diff}^\semcr(D, \kappa(\mc{Q}), t) \not = \emptyset$. \boxtheorem
\end{definition}

  Notice that every $(\semc)$-preferred cause is also an actual cause.  This follows from  Proposition \ref{pro:c&r} and the fact that prioritized repairs are also S-repairs.

   Similarly to Proposition \ref{pro:r&r},  for each $\Lambda \in \nit{Diff}^\semcr(D,\kappa(\mc{Q}), t)$, it holds that $t \in \Lambda$, $t$ is a $(\semc)$-preferred cause, and also an  actual cause for $\mc{Q}$ with S-minimal contingency set $\Lambda \smallsetminus \{t\}$. In particular, $t$'s responsibility can be defined and computed as before, but now restricting its contingency sets to those of the form $\Lambda \smallsetminus \{t\}$, with $\Lambda \in
   \nit{Diff}^\semcr(D,\kappa(\mc{Q}), t)$. In this way, a causal priority relation may affect the responsibility of a cause (with respect to the non-prioritized case).

\begin{example}\label{ex:pfcex2} \ (example \ref{ex:pfcex1} cont.) \ The following BCQ query $\mc{Q}$ is true in $D$:
\begin{eqnarray*}\exists \nit{JournalN} \ \exists \nit{Paper\#}(\nit{Author}({\sf John},&&\!\!\nit{Journal}) \ \wedge\\&& \nit{Journal}(\nit{JournalN},\nit{Paper\#},{\sf XML}));
\end{eqnarray*}
 and its associated DC $\kappa(\mc{Q})$ is $\kappa$ in (\ref{eq:kappa}).

We want to obtain the preferred causes for $\mc{Q}$ being, possibly unexpectedly, true in $D$, with the following preferences: (a) We prefer those  among the \nit{Author} tuples. \ (b) It is likely that John does have a paper in TODS. So, we prefer \nit{Author(John, TODS)} not to be the cause.

These causal priorities are in  inverse correspondence with those in the second case of Example \ref{ex:pfcex1}(b) about priorities for repairs. That is, for our causal priority
relation $\succ_{\!c}$ here, its inverse $\succ_{\!c}^r$ is $\succ'$ in Example \ref{ex:pfcex1}(b). There we had $\nit{Rep}^{\succ',\nit{go}}(D,\kappa(Q)) = \{D_1, D_2\}$, which we can use to apply
 Definition \ref{def:prefCuases}.

We  obtain  as the globally-optimal causes, i.e. as  $(\succ_{\!c},\nit{go})$-causes: \linebreak \nit{Author(John, TKDE)}, \nit{Author(TODS,32,XML)} and \nit{Author(John,TODS}), all with the same responsibility, $\frac{1}{2}$.
\boxtheorem
\end{example}

Notice that Definition \ref{def:prefCuases} can be easily extended to UBCQs. This is done, as earlier in this work, by considering the set $\Sigma$ of denial constraints associated to a UBCQ.
In the other direction, we recall that if we start with a set of DCs $\Sigma$, the corresponding UBCQ is denoted with $V^{\!\Sigma}$.

As we did in the previous sections of this work, we could  take advantage of algorithmic and complexity results about prioritized repairs \cite{chomicki12,fagin15}, to obtain complexity results
for preferred causes problems. As an example, we establish the complexity of the minimal contingency set decision problem for $(\succ_{\!c},\nit{go})${\em -preferred causes}. \ignore{Recall  in Section \ref{sec:MBDcomx} we show that the similar problem for actual causes in general is tractable.}
More precisely, for an instance $D$ and a  UBCQ $\mc{Q}$,  the {\em  minimal preference-contingency set} (decision) problem is about deciding if a set of tuples $\Gamma$ is an S-minimal contingency set
 associated to a $(\succ_{\!c},go)$-preferred cause $t$.

\ignore{
\comlb{Again, see the underlined part.}
\combabak{ Here we can get rid of it. However, we need to modify the MPCDP definition (as I did below) to deal with all contingencies }
}
\ignore{
\combabak{We have to make the proposition over UBCQs. This is because the hardness of global optimal repair checking has been proved via a set of particular FDs in \cite{chomicki12}.}
}

  \vspace{1mm} \noindent \red{{\bf \em Notation:}} \ $\nit{Cont}^\semc(D,\mc{Q},t) := \{\Lambda \smallsetminus \{t\}~|~ \Lambda \in \nit{Diff}^\semcr(D,\kappa(\mc{Q}), t)\}$ is
  the \linebreak \hspace*{1.2cm} class of all S-minimal contingency sets for a $(\semc)$-preferred cause $t$.

\begin{definition}  \label{def:Pcusp} For  a UBCQ $\mc{Q}$, the {\em minimal preference-contingency set} decision problem is about membership of:

$\mc{MPCDP}(\mc{Q}):=\{(D,\succ_{\!c}, t,\Gamma)~|~ t \in D, \Gamma \subseteq  D, \mbox{ and } \Gamma \in \nit{Cont}^{\succ_{\!c},go}(D,\mc{Q},t)\}$.\boxtheorem
\end{definition}

 From Definition \ref{def:prefCuases}, there is a close connection between $\mc{MPCDP}$ and  the {\em global-optimal repair checking problem}, i.e. about deciding if an instance $D'$ is a $(\succ,\nit{go})$-repair of $D$ with respect to a set of denial constraints. If we accept functional dependencies (FDs) among our denial constraints (and then, UBCQs that involve inequalities),
the following result can be obtained from the {\em NP}-completeness of globally-optimal repair checking \cite{chomicki12} for FDs.

\begin{proposition}\label{pro:PCSPCcpx} \em For  a UBCQ $\mc{Q}$  with inequalities,
$\mc{MPCDP}(\mc{Q})$ is {\em NP}-hard. 
\end{proposition}
\proof{It is good enough to reduce globally-optimal repair checking to our contingency checking problem. So, consider  an inconsistent instance $D$ with respect to a set of denial constraint $\Sigma$, a priority relation
for repairs $\succ$, and $D' \subseteq D$. To check if $D' \in \nit{Rep}^{\succ,go}(D,\Sigma)$ we can  check, for an arbitrary element $t \in D \smallsetminus D'$, if $(D,\succ^r, t, D \smallsetminus (D'\cup  \{t\})\in \mc{MPCDP}(V^{\!\Sigma})$. \boxtheorem\\}

It is worth contrasting this result with the tractability result in Proposition \ref{pro:CSPCcpx} for the {\em minimal contingency set decision problem} (MCSDP) for actual causes. Notice that Proposition \ref{pro:CSPCcpx}
still holds for UBCQs with inequality.

Notice that we could generalize the notion of preferred cause by appealing to any notion of repair. More precisely,  if we have a {\em repair semantics} $\sf{rSem}$ (based on tuple deletions
for DCs), we could replace $\nit{Rep}^{\sem}(D,\kappa(\mc{Q}))$ in (\ref{eq:dif}) by $\nit{Rep}^{\large \sf S}(D,\kappa(\mc{Q}))$. However, to obtain the intended results for causes, we have to be careful, as above, about a possible inverse relationship between
preference on repairs and preference on causes.

\subsection{ \ Endogenous repairs}\label{sec:endo}

The partition of a database into endogenous and exogenous tuples that is used in the causality setting may also be of interest in the context of repairs. Considering that we should
have more control on endogenous tuples than on exogenous ones, which may come from external sources, it makes sense to consider
{\em endogenous repairs}, which would be obtained by updates (of any kind) on endogenous tuples only. (Of course, a symmetric treatment of ``exogenous" repairs is also possible; what is relevant here is the partition.)

For example, in the case of DCs,
   endogenous repairs would be obtained by deleting endogenous tuples only.
More formally, given $D = D^n \cup D^x$, possibly inconsistent with a set of DCs $\Sigma$, an {\em endogenous repair} $D'$ of $D$ is a maximally consistent
sub-instance of $D$ with $D \smallsetminus D' \subseteq D^n$, i.e. $D'$ keeps all the exogenous tuples of $D$.
If endogenous repairs form the class
$\nit{Srep}^n(D,\Sigma)$, it holds   $\nit{Srep}^n(D,\Sigma) \subseteq \nit{Srep}(D,\Sigma)$.

\begin{example}\label{ex:endRep}
 \ Consider $D= D^n \cup D^x$, with
$D^n=\{R(a_2,a_1), R(a_4,a_3), S(a_3),$ $ S(a_4)\}$ and $D^x=\{ R(a_3,a_3), S(a_2)\}$, and the DC $\kappa\!: \ \neg \exists x y (S(x) \wedge R(x, y) \wedge S(y))$.

Here, $\nit{Srep}(D, \kappa)$ $=$ $\{D_1,$ $D_2,$ $D_3\}$, with
$D_1= \{R(a_2,a_1), R(a_4,a_3), $ $ R(a_3,a_3),$ $ S(a_4), S(a_2)\}$, \
$D_2 = \{ R(a_2,a_1), S(a_3),S(a_4), S(a_2)\}$, \ and \
$D_3 = \{R(a_2,a_1),$ $R(a_4,a_3),  S(a_3), S(a_2)\}$.
\ The  only endogenous S-repair is $D_1$. \boxtheorem
   \end{example}

   In this section, without trying to be exhaustive or detailed, we consider the possibility of defining endogenous repairs on the basis of a suitable priority relation $\succ$ on tuples,\footnote{Pairs of conflicting tuples would inherit the priority relationships from the general priority relation.} while
at the same time taking advantage of the \nit{op} optimality condition considered in Section \ref{sec:pref-rep}.\footnote{Of course, we could use other
optimality criteria at this points, but considering all possibilities is beyond the scope of this work.}

First, if we assume that relation $\succ'$, the extension of $\succ$, is such, that $t \succ' t'$ when $t \in D^x$  and $t' \in D^n$ ($\succ'$ is $\succ$ if the latter already has this property), then it is easy
to verify that every endogenous S-repair globally improves any non-endogenous S-repair. As a consequence, if there is an endogenous S-repair, then all the $(\succ',\nit{go})$-repairs are endogenous.  Notice that the
extension $\succ'$ may destroy the acyclicity assumption on the priority relation, because we are starting from a given (acyclic) relation $\succ$, which we are now extending.

It might be the case that there is no endogenous S-repair, in which case non-endogenous S-repairs would not the improved by an endogenous one. So, \red{if we want
to prevent the existence of non-endogenous repairs,} we can add an extra, dummy predicate $D(\cdot)$ to the schema, and the endogenous tuple $D(d)$ to $D$. We modify every DC \red{in $\Sigma$}, say \ $\kappa\!: \ \leftarrow \ C(\bar{x})$, by adding
an extra, dummy condition:  \ $\kappa^d\!: \ \leftarrow \ D(d), C(\bar{x})$, \red{obtaining a set $\Sigma^d$ of DCs}.  In this case, the S-repairs will be: \ $D^d:=D \smallsetminus \{D(d)\}$, which is endogenous, and also all those S-repairs
of $D$ with respect to $\Sigma$ (now each including $D(d)$). \red{The latter are all non-endogenous.} If we assume that \red{$t \succ' D(d)$, for every $t \in D^x$}, then every non-endogenous S-repair will be improved by $D^d$, and will not be considered.

\ignore{If we  get rid of the original priority relationships $t \succ t'$, with $t \in D^n, t' \in D^x$, if any, then the $(\succ',\nit{go})$-repairs
of $D \cup \{D(d)\}$ with respect to \red{$\kappa^d$} will be all endogenous, namely  $D^d$ plus the $D' \cup \{D(d)\}$, where $D'$ is an endogenous $(\succ,\nit{go})$- repair of $D$ with respect to $\Sigma$.
In particular, if the only priority relationships are $D(d) \succ t$, with $t \in D^n$, then we obtain as repairs: $D^d$ plus all the endogenous S-repairs of $D$ with respect to $\Sigma$ (each of them now including also the tuple
$D(d)$).}

\ignore{More generally, if we have a set $\Sigma$ of DCs, and a priority relation, $\succ$, without relationships of the form $t \succ t'$, with $t \in D^n, t' \in D^x$, then there will be
endogenous S-repairs of $D \cup \{D(d)\}$ with respect to \red{$\Sigma^d$}; and the $(\succ',\nit{go})$-repairs
of $D \cup \{D(d)\}$ with respect to \red{$\Sigma^d$} will be all endogenous, namely  $D^d$ plus the $D' \cup \{D(d)\}$, where $D'$ is an endogenous $(\succ,\nit{go})$- repair of $D$ with respect to $\Sigma$.
In particular, if the only priority relationships are $D(d) \succ t$, with $t \in D^n$, then we obtain as repairs: $D^d$ plus all the endogenous S-repairs of $D$ with respect to $\Sigma$ (each of them now including also the tuple
$D(d)$).}

\subsection{ \ Null-based causes}\label{sec:null}

Consider an instance $D = \{R(c_1, \ldots, c_n), \ldots\}$ that may be
inconsistent with respect to a set of DCs. The allowed repair updates are changes of
attribute
values by the constant \nit{null}. We assume that $\nn$ does not join with
any other value, including \nn \ itself.

In order to keep track of changes, we may introduce numbers as first
arguments in tuples, as  global tuple identifiers (ids). So, $D$ becomes $D = \{R(1;c_1, \ldots, c_n),
\ldots\}$. Assume  that $\nit{id}(t)$ returns the id of the tuple
$t \in D$. For example, $\nit{id}(R(1;c_1, \ldots, c_n)) = 1$.

If, by updating $D$ into $D'$ in this way, the value of the $i$th attribute
in $R$ is changed to \nn, then the change is captured as the string
$R[1;i]$. These strings are collected
forming the set $\nit{Diff}^\nn(D,D')$. For example, if $D = \{R(1;a,b),
S(2;c,d), S(3;e,f)\}$ is changed into $D' = \{R(1;a,\nn), S(2;\nn,d),$ $
S(3;\nn,\nn)\}$, we have $\nit{Diff}^\nn(D,D') = \{R[1;2], S[2;1],
S[3;1], S[3;2]\}$.

A \nit{null}-repair of $D$ with respect to a set of DCs $\Sigma$ is a consistent
instance $D'$, such that $\nit{Diff}^\nn(D,D')$ is minimal under set
inclusion.\footnote{An alternative, but equivalent
formulation can be found in \cite{lechen}.}
$\nit{Rep}^\nit{null}(D,\Sigma)$ denotes the class of null-based repairs of
$D$ with respect to $\Sigma$.

\begin{example}\label{ex:nullReps} \ (example \ref{ex:endRep} cont.) \
Consider the following inconsistent instance with respect to DC \ $\kappa\!: \ \neg
\exists x y (S(x) \wedge R(x, y) \wedge S(y))$:

\centerline{$D= \{R(1;a_2,a_1), R(2;a_3,a_3),R(3;a_4,a_3),$ $S(4;a_2), S(5;a_3),
S(6;a_4)\}$. }

For simplicity, we do not make any difference between endogenous and
exogenous
tuples.
Here, the  class of {\em null-based repairs}, \
$\nit{Rep}^\nit{null}(D,\kappa)$, is formed by:

$D_1= \{R(1;a_2,a_1),R(2;a_3,a_3), R(3;a_4,a_3),S(4;a_2), S(5;\nit{null})  ,
S(6;a_4)\}$,

$D_2 = \{ R(1;a_2,a_1), R(2;\nit{null},a_3), R(3;a_4,\nit{null}), S(4;a_2),
S(5;a_3),S(6;a_4)\}$,

$D_3 = \{ R(1;a_2,a_1), R(2;\nit{null},a_3), R(3;a_4,a_3), S(4;a_2),
S(5;a_3),S(6;\nn)\}$,

$D_4 = \{ R(1;a_2,a_1), R(2;a_3,\nit{null}), R(3;a_4,\nit{null}), S(4;a_2),
S(5;a_3),S(6;a_4)\}$,

$D_5 = \{ R(1;a_2,a_1), R(2;a_3,\nit{null}), R(3;\nn,a_3), S(4;a_2),
S(5;a_3),S(6;a_4)\}$,

$D_6 = \{ R(1;a_2,a_1), R(2;a_3,\nit{null}), R(3;a_4,a_3), S(4;a_2),
S(5;a_3),S(6;\nn)\}$.

\vspace{1mm}
\noindent Here,  \red{$\nit{Diff}^\nn(D,D_2) = \{ R[2;1], R[3;2]\}$, and $\nit{Diff}^\nn(D,D_3) = \{R[2;1],$ $ S[6;1]\}$}. \boxtheorem
\end{example}

According to the motivation provided at the beginning of this section,
we can now define causes appealing to the class of null-based repairs of $D$. Since
repair actions in this case, are attribute-value changes,  causes can be defined at both the tuple and attribute levels.
The same applies to the definition of responsibility (in this case generalizing Proposition \ref{pro:r&r}).

\begin{definition} \label{def:attCuases} \
 For $D$ an instance and
$\mc{Q}$ a BCQ, and $t \in D$ be a tuple of the form $R(i;c_1, \ldots, c_n)$.
   \begin{enumerate}
 \item [(a)]

 $R[i;c_j]$ is a \ {\em null-based attribute-value cause} for $\mc{Q}$ \  if there is $D' \in \nit{Rep}^\nn(D,$ $\kappa(\mc{Q}))$ with $R[i;j] \in \nit{Diff}^\nn(D,D')$.

 (That is, the value $c_j$ for attribute $A_j$ in the tuple is a cause if it is changed into a null in some repair.)

\item [(b)] $t$ is a \ {\em null-based tuple cause} for $\mc{Q}$ \  if  some $R[i;c_j]$ is a {\em null-based attribute-value cause} for $\mc{Q}$.

(That is, the whole tuple is a cause if at least one of its attribute values is changed into a null in some repair.)

\item [(c)] The responsibility, $\rho^{\mbox{\small \it t-null}}(t)$, of $t$, a \ {\em null-based tuple cause} for $\mc{Q}$, is the inverse of \
$\nit{min}\{|\nit{Diff}^\nn(D,D')|~:~R[i;j] \in  \nit{Diff}^\nn(D,D'), \mbox{ for some } j, \mbox{ and}$ $ D' \in  \nit{Rep}^\nn(D,\kappa(\mc{Q}))\}$.

\item [(d)] The responsibility, $\rho^{\mbox{\small \it a-null}}(R[i;c_j])$, of $R[i;c_j]$, a \ {\em null-based attribute-value cause} for $\mc{Q}$, is the inverse of \
$\nit{min}\{|\nit{Diff}^\nn(D,D')|~:~R[i;j] \in  \nit{Diff}^\nn(D,$ $D'),$  $\mbox{and } D' \in  \nit{Rep}^\nn(D,\kappa(\mc{Q}))\}$.
\boxtheorem
 \end{enumerate}
\end{definition}
In cases (c) and (d) we minimize over the number of changes in a repair that are made together with that of the candidate tuple/attribute-value to be a cause. In the case of a tuple cause,
any change made in one of its attributes is considered in the minimization. For this reason, the minimum may be smaller than the one for a fixed attribute
value change; and so the responsibility at the tuple level may be greater than that at the attribute level. More precisely,
 if $t = R(i;c_1, \ldots, c_n) \in D$, and $R[i;c_j])$ is a {\em null-based attribute-value cause}, then
it holds $\rho^{\mbox{\small \it a-null}}(R[i;c_j]) \leq \rho^{\mbox{\small \it t-null}}(t)$.

\begin{example} (ex. \ref{ex:nullReps} cont.) \ Consider $R(2;a_3,a_3) \in D$. Its projection on its first (non-id) attribute, $R[2;a_3]$, is an attribute-level
cause since $R[2;1] \in
\nit{Diff}^\nn(D,D_2)$. Also $R[2;1] \in
\nit{Diff}^\nn(D,D_3)$.

Since $|\nit{Diff}^\nn(D,D_2)| = |\nit{Diff}^\nn(D,D_3)| = 2$, it holds $\rho^{\mbox{\small \it a-null}}(R[2;1]) = \frac{1}{2}$.

Clearly $R(2;a_3,a_3)$ is  a {\em null-based tuple cause} for $\mc{Q}$, with $\rho^{\mbox{\small \it t-null}}(t) = \frac{1}{2}$.
\boxtheorem
\end{example}

Notice that the definition of tuple-level responsibility, i.e. case (c) in Definition \ref{def:attCuases}, does not take into account that a same id, $i$, may appear
several times in a $\nit{Diff}^\nn(D,D')$. In order to do so, we could redefine the size of the latter by taking into account those multiplicities. For example, if we decrease the size of the \nit{Diff} by one
with every repetition of the id, the
responsibility  for
a cause may (only) increase, which makes sense.

\section{ \ \red{Discussion and Conclusions}}\label{sec:disc}

Our work  opens interesting research directions, some of which are briefly discussed below. They are matter of ongoing and future research.

\subsection{ \ Endogenous repairs} \label{sec:endrep}

As discussed in Section \ref{sec:pref-cause}, the partition of a database into endogenous and exogenous tuples may also be of interest in the context of repairs.
We may prefer endogenous repairs that change (delete in this case) only endogenous tuples. However,
   if there are no endogenous tuples, a preference condition could be imposed on repairs, keeping those that change exogenous tuples the least. This is something to explore.

  As a further extension, it could be possible to assume that combinations of (only) exogenous tuples never violate the integrity constraints, which could be checked
   at upload time. In this sense, there would be a part of the database that is considered to be consistent, while the other is subject to possible repairs. For somehow
   related research, see  \cite{greco14}.


 Going a bit further, we could even consider the relations  in the database with an extra, binary  attribute, $N$, that is used to annotate if a tuple is
  endogenous or exogenous (it could be both), e.g. a tuple like $R(a,b, \nit{yes})$. integrity constraints could be annotated too, e.g. the ``exogenous" version of DC $\kappa$, could be
   $\kappa^E\!: \ \leftarrow P(x, y,\nit{yes}),R(y, z,\nit{yes})$, and could be assumed to be satisfied.

\subsection{ \ Objections to causality}
Causality as introduced by Halpern and Pearl in \cite{Halpern01,Halpern05}, aka. HP-causality, is the basis for the notion of causality in \cite{Meliou2010a}. HP-causality has been the object of some criticism \cite{halpern14}, which is justified in some (more complex, non-relational) settings, specially due to the presence of different kinds of {\em logical variables} (or lack thereof). In our context
 the objections do not apply:  variables just  say  that a certain tuple belongs to the instance (or not); and for relational databases the closed-world assumption applies.
In \cite{halpern14,halpern15}, the definition of HP-causality is slightly modified. In our setting, this modified definition does not change actual causes or their properties.


\subsection{ \ Open queries} \ We have limited our discussion to boolean queries. It is possible to extend our work
to consider conjunctive queries with free variables, e.g. $\mc{Q}(x)\!: \exists yz(R(x,y) \wedge S(y,z))$. In this case,
a query answer would be of the form $\langle a\rangle$, for $a$ a constant, and causes would be found for such an answer.
In this case, the associated DC would be of the form $\kappa^{\langle a\rangle}\!: \ \leftarrow R(a,y), S(y,z)$, and
the rest would be basically as above.


\ignore{
\paragraph{\bf Algorithms and complexity.} \ Given the connection between causes and different kinds of repairs, we might take advantage
for causality of algorithms and complexity results obtained for database repairs. This is matter of our ongoing research. In this work, apart
from providing a naive algorithm for computing repairs from causes, we have not gone into detailed algorithm or complexity issues. The results
we already have in this direction will be left for an extended version of this work.
}

\subsection{ \ ASP specification of causes} \label{sec:asps}  S-repairs can be specified by means of
   {\em answer set programs} (ASPs) \cite{tplp03,barcelo03}, and C-repairs too, with the use of weak program constraints \cite{tplp03}. This should allow for the
   introduction of ASPs in the context of causality, for specification and reasoning.
    There are also ASP-based specifications of diagnosis \cite{eiter99} that could be brought into a more complete picture.

\subsection{ \ Causes and functional dependencies, and beyond}

Functional dependencies are DCs with conjunctive violation views with
    inequality, and are still monotonic. There is much research on repairs and consistent query answering for functional dependencies, and more complex integrity constraints \cite{2011Bertossi}. In causality, mostly
    CQs without built-ins have been considered. \red{The repair connection could be exploited  to obtain more refined results for causality and CQs with inequality, and also other classes of queries, even non-monotonic ones, that correspond violation views for other kinds of integrity constraints. In a different, but related direction, causality for monotonic queries in the presence of
    integrity constraints has been investigated in \cite{flairs16}.}
    \ignore{It is possible that causality can be extended to conjunctive queries with built-ins through the repair connection; and also to non-conjunctive queries via repairs with respect to\ more complex
    integrity constraints.}

 \subsection{ \ View updates and abduction}  Abduction \cite{Console91,EiterGL97} is another form of model-based diagnosis, and is related to the subjects investigated in this work. The {\em view update problem}, about updating a database through views,
is a classical problem in databases that has been treated through abduction
\cite{Kakas90,Console95}. User knowledge imposed through view updates creates or reflects {\em uncertainty} about the base data, because alternative base instances may give an account
of the intended view updates.
The view update problem, specially in its particular form of {\em deletion propagation}, has been recently related in \cite{benny12a,benny12b} to causality as introduced in
\cite{Meliou2010a}. (Notice only tuple deletions are used with violation views and repairs associated to DCs.)

 Database repairs are also related to the view update problem.
Actually, {\em answer set programs} (ASP) for database repairs \cite{barcelo03} implicity repair the database by updating intentional,
annotated predicates (cf. Section \ref{sec:asps}).
Even more, in \cite{lechen}, in order to protect sensitive information, databases are explicitly and virtually ``repaired" through secrecy views that specify the
information that has to be kept secret. These are prioritized repairs that have been specified via ASPs. Abduction has been explicitly applied to database repairs \cite{arieli}.

The deep interrelations between causality, abductive reasoning, view updates and repairs are  the objects of our ongoing research efforts \cite{buda14,uai15}.\\


To conclude, let us emphasize that in this research we have unveiled and formalized some first interesting  relationships between causality in databases, database repairs, and consistency-based diagnosis. These connections allow us to apply results and techniques developed for
each of them to the others. This is  particularly beneficial for causality in databases, where still a  limited number of results and techniques have been obtained or developed.

The connections we established here inspired complexity results for causality, e.g. Theorems \ref{the:r&diag} and \ref{the:cqa&ca&cox}, and were used to prove them.
We appealed to several non-trivial results found in \cite{icdt07} (and the proofs thereof found in \cite{icdt07ext}) about repairs and CQA.
It is also the case that the
well-established hitting-set approach to diagnosis inspired
a similar  approach to causal responsibility, which in its turn allowed
us to obtain results about its fixed-parameter tractability.   It is also the case that diagnostic reasoning, as a form of non-monotonic reasoning, can provide a  solid foundation  for causality in databases and query answer explanation, in general \cite{Cheney09b,Cheney11}.

In ongoing research we have established connections between query answer causality, abductive diagnosis and database updates through views \cite{uai15}. It is interesting that several
of these areas of data management and knowledge representation, including those considered in this work, fall under what has been called ``reverse data management" tasks \cite{suciuRevDM}. Our work
establishes formal connections between them and sets the ground for further investigation into their interrelationships.

\vspace{1mm}
\noindent {\bf Acknowledgments:} \ Research funded by NSERC Discovery, and
the NSERC Strategic Network on Business Intelligence (BIN).  Conversations  with Alexandra Meliou during Leo Bertossi's visit to U. of Washington in 2011 are much appreciated.
He is also grateful to Dan Suciu and  Wolfgang Gatterbauer for their hospitality. L. Bertossi is grateful to Benny Kimelfeld for stimulating conversations.
Part of the research was developed by L. Bertossi during partial sabbatical stays at {\em LogicBlox} and {\em The Center for Semantic Web Research} (Chile). Their support is much appreciated. We appreciate the comments from the anonymous reviewers.

\end{document}